\documentclass[twocolumn,showpacs,preprintnumbers,amsmath,amsthm,amssymb]{revtex4}
\usepackage{epsfig}

\usepackage{graphicx}% Include figure files
\usepackage{dcolumn}% Align table columns on decimal point
\usepackage{bm}% bold math
\usepackage{paralist}
\usepackage{amsmath, amsthm, amssymb}
\usepackage{bbm,bm}

\newtheorem{thm}{Theorem}[section]

\newtheorem{lem}[thm]{Lemma}
\newtheorem{defn}{Definition}[section]

\begin{document}

%\preprint{APS/123-QED}

\title{Decoherence Control in Open Quantum System via Classical Feedback}% Force line breaks with \\

\author{Narayan Ganesan}
 \email{ng@ese.wustl.edu}
\author{Tzyh-Jong Tarn}%
 \email{tarn@wuauto.wustl.edu}
\affiliation{%
Electrical and Systems Engineering.\\
Washington University in St. Louis
}%


\begin{abstract}
In this work we propose a novel strategy using techniques from systems theory to completely eliminate decoherence
and also provide conditions under which it can be done so. A novel construction employing an auxiliary system, the
bait, which is instrumental to decoupling the system from the environment is presented. Our approach to
decoherence control in contrast to other approaches in the literature involves the bilinear input affine model of
quantum control system which lends itself to various techniques from classical control theory, but with
non-trivial modifications to the quantum regime. The elegance of this approach yields interesting results on open
loop decouplability and Decoherence Free Subspaces(DFS). Additionally, the feedback control of decoherence may be
related to disturbance decoupling for classical input affine systems, which entails careful application of the
methods by avoiding all the quantum mechanical pitfalls. In the process of calculating a suitable feedback the
system has to be restructured due to its tensorial nature of interaction with the environment, which is unique to
quantum systems. The results obtained are qualitatively different and superior to the ones obtained via master
equations. Finally, a methodology to synthesize feedback parameters itself is given, that technology permitting,
could be implemented for practical 2-qubit systems to perform decoherence free Quantum Computing.
\end{abstract}

\pacs{}% PACS, the Physics and Astronomy
\maketitle

\section{Introduction}
Various authors have studied control of decoherence of an open quantum system. Decoherence Free Subspaces(DFS)
help preserve quantum information in an open quantum system. However, the presence of symmetry breaking
perturbations or control hamiltonians acting on an open quantum system which is essential to performing arbitrary
transforms in the system hilbert space $\mathcal{H}_s$, could also lead to loss of information by inevitable
transfer of states out of DFS, due to the nature of the control hamiltonians. Hence this renders the quantum
system at best a noiseless memory, much less a dynamic quantum computer, whose state needs to be transformed in
order to perform computations. Recently Lidar and Wu \cite{lidar2},\cite{lidar3}, Kielpinski et.
al.\cite{kielpinski}, Brown et. al \cite{brown} have proposed a combination of open loop bang-bang pulses,
universal control in order to perform computation within the DFS via control pulses. In this work we propose a
novel strategy, exploiting the geometry of the bilinear control system on the analytic manifold to {\it completely
eliminate} decoherence in the presence of symmetry breaking control hamiltonians and still preserve complete
controllability of the system in order to perform arbitrary transforms. We also explore the possibilities and
provide conditions under which it can be done so. This {\it unified} approach to control of decoherence lets us
analyze the open loop decoupling problem which directly leads us to the existence of DFS and secondly closed loop
decoupling via a classical feedback to the control system which leads us to robust decoherence control. This work
is a continuation of the previous results\cite{ganesan} wherein some of the theoretical groundwork was laid to
study the problem of open loop decoupling, which are now extended to closed loop control and feedback design here.
The approach used here is fundamentally different from approaches adopted by other authors in that $(i)$ the
bilinear form of control system is used which is amenable to classical systems theoretical results instead of the
stochastic master equation for the state evolution, $(ii)$ the approach does not aim at mitigating or slowing down
the decoherence rate rather aims at completely eliminating via a suitable non-linear feedback. The experimental
feasibility is discussed for a finite state environment acting on a two qubit system which is a rather reasonable
approximation. A procedure to compute the feedback using the invariant subspace for a system is provided. A
detailed step by step algorithm to determine the {\it invariant subspace} itself on the tangent space
$T_\xi(\mathcal{M})$ is also provided. In order to compute the feedback parameters a good estimate of state of the
system is essential. A reliable information extraction scheme utilizing indirect continuous measurement via a
quantum probe in the context of a decohering quantum system was studied in\cite{ganesan1}.

\section{Previous Work}
Consider an open quantum system interacting with the environment described by,
\begin{eqnarray*}
&\frac{\partial\xi(t,x)}{\partial t}=&[H_0\otimes \mathcal{I}_e(t,x)+\mathcal{I}_s \otimes H_e(t,x)+H_{SE}(t,x)
\\&& +\sum_{i=1}^{r}u_i(t)H_i\otimes \mathcal{I}_e(t,x)]\xi(t,x) \label{opqusys}
\end{eqnarray*}
Here the argument $x$ denotes the spatial dependance of the combined system-environment state $\xi(t,x)$ as well
as control hamiltonians $H_i$, and where $u_i$ are the strength of the control respectively. $H_0, H_E, H_{SE}$
are the system, environment and interaction hamiltonian acting on $\mathcal{H}_s$, $\mathcal{H}_e$ and
$\mathcal{H}_s\otimes\mathcal{H}_e$ (system, environment and the combined) Hilbert spaces respectively. For ease
of notation we will suppress the spatial dependance. Define an output equation which could either be a
non-demolition measurement or a general bilinear form given by,
\begin{equation}
y(t)=\langle \xi(t)|C(t)|\xi(t)\rangle \label{output}
\end{equation}
where again $C(t,x)$ is assumed to be time-varying operator acting on the system Hilbert space. For instance for a
finite system the non-hermitian operator $C=|m\rangle\langle n|$ when plugged in eq. (\ref{output}) would yield
the coherence between the respective states of the system or for an electro-optic system the operator
$C=a\exp(i\omega t)+a^\dagger\exp(-i\omega t)$ would yield the output of a real non-demolition observation
performed on the system. In order to study the invariance properties with respect to the system dynamics of the
above time dependent quantum system, we define $f(t,x,u_1,\cdots,u_r,H_{SB})=y(t,\xi) \mbox{ for } t\in[t_0,t_f]$
to be a complex scalar map as a function of the control functions and the interaction Hamiltonian $H_{SB}$ over a
prescribed time interval. The function $f$ is said to be invariant or the signal $y(t,\xi)$ is said to decoupled
from the interaction Hamiltonian $H_{SB}$ if,
\begin{equation}
f(t,x,u_1,\cdots,u_r,H_{SB}) = f(t,x,u_1,\cdots,u_r,0) \label{cond9}
\end{equation}
for all admissible control functions $u_1,\cdots,u_r$ and a given interaction Hamiltonian $H_{SB}$. Then the
condition for such an output signal to be decoupled from the interaction hamiltonian in the open loop case is
given by the following theorem\cite{ganesan}, which follows an iterative construction in terms of system
operators.

%The necessary and sufficient conditions for openloop decouplability and the necessary conditions for
%decouplability via feedback can now be recalled as discussed in the previous chapter. Let $x_1=t$, the new
%equation governing the evolution of the time-varying system can be written as,
%\begin{align}
%\frac{\partial}{\partial t} \left(\begin{array}{c}x_1 \\
%\xi(t,x)\end{array}\right) &= \left( \begin{array}{c} 1 \\
%(H_0(x_1,x)+H_P(x_1,x)+ H_E(x_1,x))\xi(t,x)
%\end{array} \right) \nonumber \\
%&+ \left( \begin{array}{c} 0 \\ u_i H_i(x_1,x)\xi(t,x) \end{array} \right)+
%\left( \begin{array}{c} 0 \\ g H_{SP}(x_1,x)\xi(t,x) \end{array} \right)\label{change_the_label} \\
%&+ \left( \begin{array}{c} 0 \\ H_{SE}(x_1,x)\xi(t,x) \end{array} \right) \nonumber
%\end{align}
%with the output
%\begin{equation}
%y(t,\xi)=\langle \xi(t,x) | C(t,x) |\xi(t,x) \rangle \label{opeq}
%\end{equation}
%and the output equation given by,

The vector fields $K_0=\left(\begin{array}{c}1\\(H_0+H_e)\xi(x,t)\end{array}\right),\\
K_i=\left(\begin{array}{c}0\\ H_i\xi(x,t)\end{array}\right)$, $K_p=\left(\begin{array}{c}0\\
H_{SP}\xi(x,t)\end{array}\right)$ and $K_I=\left(\begin{array}{c}0\\H_{SB}\xi(x,t)\end{array}\right)$
corresponding to drift, control and interaction can be identified to contribute to the dynamical evolution. It was
already noted that the the system was said to be decoupled if it satisfied equations \ref{invsp}, namely,
\begin{align}
L_{K_I} y(t,\xi) &= 0 \nonumber \\
L_{K_I} L_{K_{i_0}} \cdots L_{K_{i_n}} y(t,\xi) &= 0 \label{invsp}
\end{align}
Recalling,
\begin{thm}\label{thm1}
Let
\begin{align*}
&\mathcal{C}_0=C(t)\\
&\vdots\\
&\tilde{C}_n=\mbox{span}\{ad^j_{H_i}\mathcal{C}_{n-1}(t)|j=0,1,\ldots;i=1,\ldots,r\}\\
&\mathcal{C}_n=\left\{ \left(ad_H+\frac{\partial}{\partial t}\right)^j\tilde{C}_n; j=0,1,\cdots \right\}\\
%&\mathcal{C}_n=ad_H\tilde{C}_n+\frac{\partial \tilde{C}_n}{\partial t}\\
&\vdots
\end{align*}
\end{thm}
Define a distribution of quantum operators,
$\tilde{\mathcal{C}}(t)=\Delta\{\mathcal{C}_1(t),\mathcal{C}_2(t),\cdots{},\mathcal{C}_n(t),\cdots{}\}$. The
output equation (\ref{opeq}) of the quantum system is decoupled from the environmental interactions if and only
if, \\
Case (I): Open Loop,
\begin{equation}
[\tilde{\mathcal{C}}(t), H_{SB}(t)]=0 \label{ic}
\end{equation}
Case (II): Whereas the {\it necessary} conditions for Closed Loop control is,
\begin{align*}
&[C,H_{SB}]=0\\
&[\tilde{\mathcal{C}}(t), H_{SB}(t)]\subset \tilde{\mathcal{C}}(t)
\end{align*}
In this work we will be primarily concerned with designing feedback for quantum systems of the form
$u=\alpha(\xi)+\beta(\xi)v$ where $\alpha$ and $\beta$ are real vector and a full rank real matrix of the state
(or its estimate thereof) of dimension $1\times r$ and $r\times r$ respectively. We examine a few systems of
interest with control hamiltonians, that might be decoupled via feedback of the above form.
\begin{defn}\label{obs}
The vector field $K_\tau$ satisfying equations (\ref{invsp}) is said to be in the orthogonal subspace of the
observation space spanned by the one-forms
\begin{align}
&dy(t,\xi),dL_{K_{i_0}}y(t,\xi),\cdots, dL_{K_{i_0}} \cdots L_{K_{i_n}} y(t,\xi), \cdots\\
&\forall 0\leq i_0,\cdots, i_n \leq r \mbox{ and } n\geq 0 \nonumber
\end{align}
Denoted by $K_\tau \in \mathcal{O}^\perp$
\end{defn}
\begin{lem}
The distribution $\mathcal{O}^\perp$ is invariant with respect to the vector fields $K_0,\cdots,K_r$ under the Lie
bracket operation. (i.e) if $K_\tau \in \mathcal{O}^\perp$, then $[K_\tau,K_i] \in \mathcal{O}^\perp$ for
$i=0,\cdots, r$ \end{lem}

\section{A Single Qubit System}
Consider a single qubit spin-1/2 system coupled to a bath of infinite harmonic oscillators through an interaction
hamiltonian $H_{SB}$. The hamiltonian of the system+bath can be written as,
\[
H=\frac{\omega_0}{2}\sigma_z + \sum_k \omega_k b_k^\dagger b_k + \sum_k \sigma_z(g_k b_k^\dagger + g_k^* b_k)
\]
where the system is acted upon by the free hamiltonian $H_0$ and the decoherence hamiltonian $H_{SB}$. As is well
known there is a rapid destruction of coherence between $|0\rangle$ and $|1\rangle$ according to the decoherence
function given by \cite{proto}. In order to cast the above problem in the present framework we consider a bilinear
form of an operator $C$ that monitors coherence between the basis states. Considering $C$ to be the non-hermitian
operator $|0\rangle\langle 1|$ we have a function $y(t)$ given by $y(t)=\langle \xi(t)|C|\xi(t)\rangle$ that
monitors coherence between the states $|0\rangle$ and $|1\rangle$. The problem now reduces to analyzing the
applicability of the theorem \ref{thm1} to the given system. It can be seen right away that the condition
$[\tilde{\mathcal{C}}, H_{SB}] \neq 0$ for the distribution $\tilde{\mathcal{C}}$ defined previously, as
calculated in the previously\cite{ganesan}. This implies that the coherence is not preserved under free dynamics
or in presence of open loop control. In order to eliminate this decoherence by feedback we now assume the system
to be acted upon by suitable control hamiltonians $\{H_1,\cdots, H_r\}$ and corresponding control functions
$\{u_1,\cdots, u_r\}$. As we pointed out earlier the necessary condition is relaxed to $[\tilde{\mathcal{C}},
H_{SB}] \subset \tilde{\mathcal{C}}$, with the operators $C$ and $H_{SB}$ still required to commute with each
other $[C,H_{SB}]=0$. For the single qubit example the second condition fails to hold, again as
outlined\cite{ganesan}, thus leaving the system unable to be {\it completely} decoupled and hence vulnerable to
decoherence even in the presence of closed loop and feedback control.
\begin{figure}
\begin{center}
\epsfysize=1.55in \epsfxsize=3.0in \epsffile{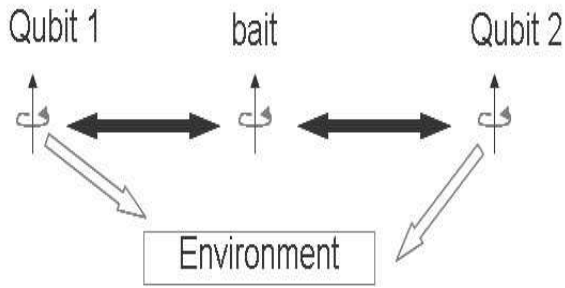}
\end{center}
\caption{The 2 Qubit system is allowed to interact with another qubit, the {\it bait} whose interaction with the
thermal bath is controlled.}
\end{figure}
\section{Two Qubit Case}
In case of two or multiple qubits there always exist Decoherence Free Subspaces (DFS) that are immune to the
decohering hamiltonian. Recently, Fortunato et. al\cite{fortunato}, Mohseni et. al\cite{mohseni}, Ollerenshaw et.
al\cite{ollerenshaw} proposed and demonstrated computation within the DFS. However it is not certain that the
system could be contained within the DFS at all times under the action of the control hamiltonians $\sigma_x,
\sigma_y$ for the system. With the effort to steer within the DFS the authors of above work could show an
improvement to previous methods, but still prone to effects of decoherence. As a simple calculation suggests that
with the initial state $c_1|01\rangle+c_2|10\rangle$, within the DFS for a 2-qubit system and after a time $t$ of
control acting on the first qubit which transforms the state to $c_1(\cos t|0\rangle +\sin t|1\rangle)|1\rangle +
c_2(\cos t|1\rangle +\sin t|0\rangle)|0\rangle$ which is clearly out of the DFS. Recently Lidar and Wu
\cite{lidar2},\cite{lidar3}, Kielpinski et. al.\cite{kielpinski}, Brown et. al \cite{brown} have proposed a
combination of open loop bang-bang pulses, universal control and DFS in the context of ion trap quantum computers
to perform computation within the DFS via control pulses, which again produces an improvement over previous
results but still prone to decohering effects. However we follow a different control strategy where in we seek to
completely eliminate the influence of $H_{SB}$ based on feedback control and a novel construction in order to
perform decoherence free control. The corresponding 2-qubit control system can be written as,
\begin{align}
&\frac{\partial |\xi(t)\rangle}{\partial t} &=& \left( \sum_{j=1}^2 \frac{\omega_0}{2}\sigma_z^{(j)} + \sum_k
\omega_k b_k^\dagger b_k\right)|\xi(t)\rangle \label{2qubitcs}\\
&&&+ \sum_k \left(\sum_j \sigma_z^{(j)}\right)(g_k b_k^\dagger + g_k^* b_k) |\xi(t)\rangle \nonumber \\
&+ (u_1(t) \sigma_x^{(1)} &+& u_2(t) \sigma_y^{(1)} + u_3(t) \sigma_x^{(2)} + u_4(t) \sigma_y^{(2)})|\xi(t)\rangle
\nonumber
\end{align}

which satisfies the basic necessary condition $[C,H_{SB}]=0$ but not the stronger condition provided in Case$(ii)$
of the theorem. Hence the system would eventually leave the DFS and is susceptible to decoherence in the presence
of arbitrary control, in other words, not entirely decoupled from $H_{SB}$. In order to analyze the system and the
conditions in the presence of a classical state feedback $u=\alpha(\xi(t))+\beta(\xi(t)).v$ the corresponding
conditions $(ii)$ of the theorem are to be examined. Since the operator $H_{SB}\in\mathcal{B}(\mathcal{H}_s\otimes
\mathcal{H}_e)$, the set of skew hermitian linear operators acting non-trivially on both system and environment
hilbert space, whereas the operators in the distribution $\tilde{\mathcal{C}}$ for the above control system is
confined to $\mathcal{B}(\mathcal{H}_s)$ that act trivially on the environment hilbert space. Hence the necessary
condition specified in Theorem\ref{thm1} would not be satisfied non-trivially unless the distribution
$\tilde{\mathcal{C}}$ acted non-trivially on both $\mathcal{H}_s$ and $\mathcal{H}_e$. In other words the
distribution includes operators of the form $\sum A_\alpha\otimes B_\alpha$ for a countable index set $\{\alpha\}$
and operators $A_\alpha$ and $B_\alpha$ operating on system and environment respectively. The above forms cannot
be achieved by control hamiltonians acting only on the system. However the situation can be salvaged if one
considered a "bait" qubit whose rate of decoherence or the environmental interaction can be modulated externally
at will and the {\it bait} qubit is now allowed to interact with our qubits of interest through an Ising type
coupling. With the help of the following construction we will be able to generate vector fields of the form $K_I$
artificially, which will be seen to provide great advantage. With the coherence functional $y(t) = \langle
\xi(t)|01\rangle\langle 10|\xi(t)\rangle$ where $|\xi(t)\rangle$, the state vector is now the total wave function
of system+{\it bait}+ environment. Both the qubit systems are assumed to interact with the same environment with
the additional requirement that the bait qubit's decoherence rate be controllable. Physically this amounts to a
coherent qubit with controllable environmental interaction. The scalability and advantages of this construction
are analyzed in the next section.

The Schr\"{o}dinger equation for the above system can now be written as,
\begin{widetext}
\begin{align}
&i\hbar \frac{\partial |\xi(t)\rangle}{\partial t} = \left( \sum_{j=1}^2 \frac{\omega_0}{2}\sigma_z^{(j)} + \sum_k
\omega_k b_k^\dagger b_k\right)|\xi(t)\rangle + \sum_k \left(\sum_j \sigma_z^{(j)}\right)(g_k b_k^\dagger + g_k^*
b_k) |\xi(t)\rangle + \left(u_1(t) \sigma_x^{(1)} + u_2(t) \sigma_y^{(1)} \right.\label{baitsysSE}\\
&\left. + u_3(t) \sigma_x^{(2)} + u_4(t)\sigma_y^{(2)} + \frac{\omega_0}{2}\sigma_z^{(b)} + u_5\sigma_x^{(b)} +
u_6\sigma_y^{(b)}+u_7 J_1 \sigma_z^{(1)}\sigma_z^{(b)} + u_8 J_2 \sigma_z^{(2)}\sigma_z^{(b)} + u_9 \sum_k
\sigma_z^{(b)}(w_k b_k^\dagger + w_k^* b_k)\right)\xi(t)\rangle \nonumber
\end{align}
\end{widetext}
where $\sigma_x,\sigma_y,\sigma_z$ are regular hermitian operators and $u_1(t)$ to $u_9(t)$ are time-dependent
piecewise constant control functions. The terms of controls $u_7$ and $u_8$ are generated by the Ising type
coupling between qubits 1, 2 and the bait with the corresponding coupling constants $J_1$ and $J_2$ respectively.
The last term in the above control system is due to the interaction of the bait qubit with the environment whose
interaction enters the system in a controllable way, hence can be treated as a separate control hamiltonian.
Keeping in mind the following commutation relations between different pairs of operators,
\begin{align*}
[\sigma_x,\sigma_y]=2i\sigma_z &&[\sigma_y,\sigma_z]=2i\sigma_x &&[\sigma_z,\sigma_x]=2i\sigma_y\\
[b_k,b^\dagger_{k'}] = \delta_{kk'} &&[b_k,b^\dagger_{k}b_k] = b_k &&[b^\dagger_{k},b^\dagger_{k}b_k] =
-b^\dagger_{k}
\end{align*}
\begin{align*}
&\sigma_z = |1\rangle\langle 1| - |0\rangle\langle 0|,\sigma_x = |0\rangle\langle 1| + |1\rangle\langle 0|\\
&\sigma_y = i|0\rangle\langle 1| - i|1\rangle\langle 0|
\end{align*}
and $C=|01\rangle\langle 10|=(\sigma_x^{(1)}-i\sigma_y^{(1)})\otimes(\sigma_x^{(2)}+i\sigma_y^{(2)})/4$, we have
$[C, H_{SB}] = 0 $ and $[\tilde{\mathcal{C}},H_{SB}]$ for instance contains terms of the form $\sigma_x\otimes
I^{(2)}\otimes \sum (g_k b_k^\dagger + g_k^* b_k)$ which are not zero. Fortunately with the above construction
these terms can be seen to be present in the distribution $\tilde{\mathcal{C}}$, which can obtained under the
sequence of operations $[C,H_1]=c_1 \sigma_z\otimes\sigma_y, [[C,H_1],H_5]=c_2 \sigma_z\otimes\sigma_x\otimes \sum
(g_k b_k^\dagger + g_k^* b_k), [[[C,H_1],H_5],H_2]= c_3 \sigma_x\otimes I^{(2)}\otimes \sum (g_k b_k^\dagger +
g_k^* b_k)$ and the corresponding $\sigma_y$ term is obtained via the sequence, $[[[C,H_2],H_5],H_1]$. Since both
terms are present in $\tilde{\mathcal{C}}$, so is their linear combination. Hence both the necessary conditions as
outlined by the theorem for closed loop decouplability are satisfied for the above system. Hence we are one step
closer to decoupling the coherence between the qubits from $H_{SB}$. In fact it can be seen that the operator
$H_{SB}$ itself can be generated by the control hamiltonians through the lie bracket operation $H_{SB} =
[[H_5,H_2],H_1]$ or $[[H_5,H_1],H_2]$. Hence any term in $[\tilde{\mathcal{C}},H_{SB}]$ is trivially contained in
$\tilde{\mathcal{C}}$. Hence, it might seem at first that the effects of $H_{SB}$ on the system could be nullified
by {\em generating} an equivalent $-H_{SB}$ through control hamiltonians alone. But in order to generate such a
vector field one has to know before hand and as time progresses the exact values of the environmental coupling
coefficients $g_k$ which at best could only be described by a stochastic process. Hence in the light of the
aforementioned difficulty, just rendering the coherence independent of $H_{SB}$ seems like a much better
alternative.

\section{Scalability}
It can also be seen that the above approach works for finite number of qubits coupled to only one bait qubit
through the same $\sigma_x^{(i)} \sigma_y^{(j)}$ interactions. Such an interaction can be implemented using the
same technology necessary for multi-qubit quantum computers wherein a finite number of qubits are entangled to a
single qubit that is capable of readout and storage of an oracle's query results. With the underlying theory of
disturbance decoupling in place all that remains now is synthesis of the feedback control itself. Since the
conditions $[\tilde{C},H_{SB}]\subset\tilde{C}$ and $[C,H_{SB}]=0$ turn out to be necessary conditions, with the
proof of sufficiency requiring further insight into design and construction of appropriate control fields we will
for the next few sections follow an alternative formalism called an {\it Invariant Subspace} which is a part of
the tangent space $T_\xi(\mathcal{M})$ of the analytic manifold. It will be seen later that the two seemingly
different approaches viz. {\it (i)} the conditions in terms of operators of the system and {\it (ii)} The tangent
space formalism, complement one another in terms of obtaining a complete solution to the problem of disturbance
decoupling.

\section{Invariant Subspace Formalism}
Consider the necessary and sufficient conditions for decouplability
\begin{align}
L_{K_I}y(t)=0\\
L_{K_I}L_{K_0}y(t)=0
\end{align}
Hence $L_{K_0}L_{K_I}y(t)=0$. The above equations after subtraction imply $L_{[K_0,K_i]}y(t)=0$. The other
necessary conditions viz. $L_{[K_0,K_i]}L_{K_j}y(t)=0$ and $L_{K_j}L_{[K_0,K_i]}y(t)=0$ imply that
$L_{[[K_0,K_i],K_j]}y(t)=0$. In fact the above pattern of equations could be extended to any number of finite lie
brackets to conclude that
\begin{equation}
L_{[[\cdots[K_0,K_{i_1}],K_{i_2}]\cdots K_{i_k}]}y(t)=0
\end{equation}
which leads us to define a set of vector fields or distribution $\Delta$ that share the same property,
\begin{equation}
K_{\nu}\in \Delta \mbox{ s.t } L_{K_\nu}y(t)=0
\end{equation}
It is observed immediately that $K_I \in \Delta$. Such a distribution $\Delta$ is said to belong to null space of
the function $y(\xi,t)$. And from the necessary conditions listed above the distribution is observed to be
invariant under the control and drift vector fields $K_0,\cdots, K_m$, (i.e) $\forall K_{\nu}\in \Delta$,
\[
[K_\nu,K_i]\in \Delta,\forall i\in{0,\cdots,m}
\]
Simply stated,
\begin{equation}\label{}
    [\Delta, K_i]\subset \Delta, \forall i\in{0,\cdots,m}
\end{equation}
\begin{figure}
\begin{center}
\epsfysize=1.9in \epsfxsize=3.1in \epsffile{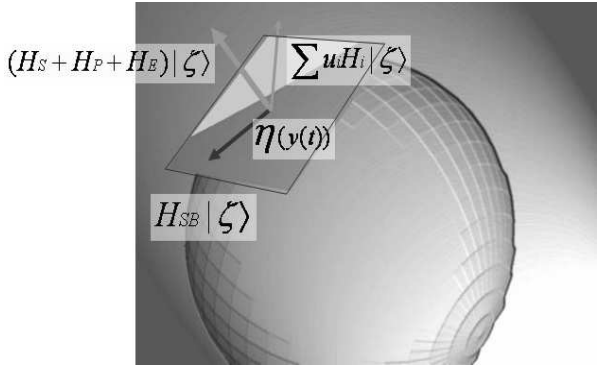}
\end{center}
\caption{The isobar of $y(t)$ is represented by the sphere and the nullspace $\mbox{ker}(dy(t)$ is a tangent to
the sphere at the point $\xi(t)$.}
\end{figure}

We will henceforth refer to the distribution as the {\it invariant distribution}. It is also to be noted that the
above calculations are reversible and the original necessary and sufficient conditions can be derived starting
from the invariant distribution. Hence the necessary and sufficient conditions for open loop decouplability can
now be restated in terms of the invariant distribution.
\begin{thm}
The output $y(t)$ is unaffected by the interaction vector field $K_I$ if and only if there exists a distribution
$\Delta$ with the following properties,\\
(i) $\Delta$ is invariant under the vector fields $K_0,K_1,\cdots, K_m$\\
(ii) $K_I\in \Delta \subset \mbox{ker}(dy(t))$
\end{thm}

Hence existence of the invariant subspace is essential to decouplability of the system in question. It is now all
the more important to determine the invariant subspace ({\it if any}) for the given system and output equation. In
order to compute the invariant distribution it properties discussed above comes in handy and provides a means to
go about computing the distribution as well.

The procedure starts out by assigning the entire null space $\mbox{ker}(dy(t))$ to invariant distribution $\Delta$
and successively removing parts of the distribution that don't satisfy the other properties (i.e), invariance with
respect the vector fields $K_0,\cdots,K_r$. In other words, remove parts of $\Delta$ whose lie brackets with
$K_0,\cdots, K_r$ do not lie within $\Delta$. Of course, the above mentioned procedure involves computing inverse
image of Lie brackets as described below.

\subsection{Invariant Distribution Algorithm}
Algorithm 1:\\
\noindent
{\it Step 1:} Let $\Delta_0=\mbox{ker}(dy(t,\xi))$.\\
{\it Step 2:} $\Delta_{i+1}=\Delta_i-\{\delta\in\Delta_i:[\delta,K_j]\notin\Delta_i\,0\leq j\leq r\}$\\
{\it Step 3:} Maximal invariant distribution is such that $\Delta^*=\Delta_i$ when $\Delta_i=\Delta_{i+1},\forall i$ .\\

The above is an iterative procedure that computes distributions $\Delta_i$ in order to arrive at the final
invariant distribution $\Delta^*=\Delta$. Where the $'-'$ is the set removal operation. Let us redefine the set to
be removed as,
\[
\mathcal{S}_i=\{\delta\in \Delta_i:[\delta,K_j]\notin \Delta_i,\forall 0\leq j \leq r\}
\]
Hence the set $\mathcal{S}_i$ can also be written as,
\begin{equation}
\mathcal{S}_i=\mbox{inv}([\Delta_i,K_j]-\Delta_i),\forall 0\leq j \leq r \label{remove_s}
\end{equation}
where 'inv' is the set theoretical inverse mapping of the linear map $[.,K_j],\forall 0\leq j \leq r$, taking
values in $\Delta$ i.e,
\[
\mbox{inv}(\tau)=\{\delta\in \Delta_i:[\delta,K_j]=\tau,\forall 0\leq j\leq r\}
\]

\begin{figure}
\begin{center}
\epsfysize=2.6in \epsfxsize=3.2in \epsffile{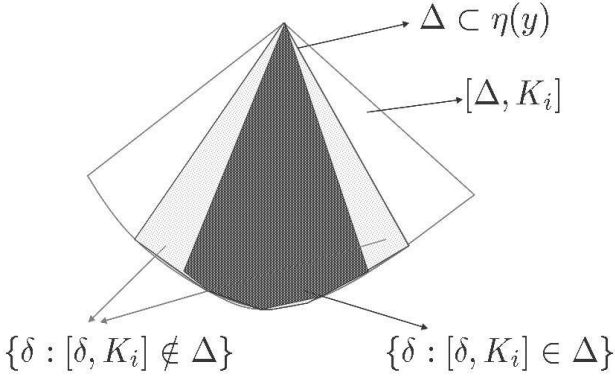}
\end{center}
\caption{Shaded portions (light and dark) mark the original distribution $\Delta_i\subset \mbox{ker}(dy(t))$. The
dark shaded portion represents the core of the distribution that is invariant and the light shaded portion, the
part of distribution that is not invariant and the white portion, image of $[.,K_i]$ that lies outside
$\Delta_i$.} \label{invalgfig}
\end{figure}

Figure(\ref{invalgfig}) outlines the schematic of the algorithm.

%******************************************************************************
%One of the foremost issues to be addressed is the convergence of the algorithm.\\
%convergence proof:\\
%*******************************************************************************

One of the foremost issues to be addressed is the convergence of the algorithm. However at this point we are not
fully equipped to study the converge as the proof below will introduce additional ideas to discuss convergence. It
is to be noted here that $\Delta_i$ is always a distribution(a vector space) for all $i$. Hence the set
$\mathcal{S}_i$ is such that, the removal of $\mathcal{S}_i$ from $\Delta_i$ results in a distribution of lower
dimension $\Delta_{i+1}$. Hence removal of the set $\mathcal{S}_i$ removes a subspace $\tilde{\Delta}_i$ contained
within the distribution $\Delta_i$. Hence we have $\Delta_{i+i}\subset \Delta_{i}$ and
\[
\Delta_{i+1}+\tilde{\Delta}_i=\Delta_i
\]
Where the '+' now denotes the direct sum of two subspaces. However, The procedure outlined above is not convenient
as it involves solving for inverse mapping under lie bracket operation \ref{remove_s}. It is for this reason that
we would like to perform the calculations in the orthogonal complement within the dual space $T^*_\xi(M)$ of the
tangent space. The algorithm can now be reformulated entirely in terms of the orthogonal complement,
$\Omega_i\subset T^*_\xi(M)$ of the distribution $\Delta_i$. (i.e) the inner product,
\[
\langle \omega, \delta \rangle = 0, \forall \omega \in \Omega \mbox{ and } \delta \in \Delta
\]
denoted by $\langle \Omega, \Delta \rangle = 0$ or $\Omega = \Delta^\perp$. Hence the algorithm now starts out by
setting  $\Omega_0 = \mbox{span}(dy(t))$ and iteratively {\it adding} the subspace that was removed by the
previous removal operation and finally inverting the co-distribution so obtained to recover $\Delta$, (i.e)\\
\noindent
{\it Step 1:} Set $\Omega_0=\mbox{span}(dy(t,\xi))$.\\
{\it Step 2:} $\Omega_{i+1}=\Omega_i+{(\tilde{\Delta}_i)}^*$.\\
{\it Step 3:} The Algorithm converges to $\Omega^*=\Omega_i$ when $\Omega_{i+1}=\Omega_i,\forall i$.\\
where $(.)^*$(not to be confused with $\Delta^*$) stands for the corresponding dual vectors within the dual space
$T^*_\xi(\mathcal{M})$, (i.e), if $\tilde{\Delta}_i=\mbox{span}\{\delta_1,\cdots \delta_k\}$ then \\
${(\tilde{\Delta}_i)}^*=\mbox{span}\{\omega_1,\cdots,\omega_k\}$ where $\langle \omega_i,\delta_i\rangle=1$. Now
the task at is to determine the subspace ${(\tilde{\Delta}_i)}^*$. It is helpful to examine the relations between
the distributions $\Delta_i,\tilde{\Delta}_i$ and $\Delta_{i+1}$. Note that, $\mbox{dim}(\Delta_i) +
\mbox{dim}(\Omega_i)=N$ and $\Delta_{i+1}$ is orthogonal to $\tilde{\Delta}_i$. In fact it can also to be seen
that $\Delta_{i+1}$ is precisely,
\begin{equation}
\Delta_{i+1}=\{\delta\in\Delta_i:[\delta,K_j]\in \Delta_i,\forall 0\leq j\leq r\} \label{di1}
\end{equation}
which is a restatement of {\it Step 2:} of Algorithm 1. Hence in order to locate the subspace $\tilde{\Delta}_i$
we have to determine the complementary subspace(look for vectors that are orthogonal) to eq.(\ref{di1}) and within
$\Delta_i$. From the identities of Lie derivatives,
\begin{equation}
L_{K_j}\langle \omega, \delta \rangle = \langle L_{K_j} \omega, \delta \rangle + \langle \omega,
[\delta,K_i]\rangle \label{liedr}
\end{equation}
Hence for $\omega\in\Omega_i$ and $\delta\in\Delta_{i+1}\subset\Delta_i$, we have,$\langle \omega, \delta
\rangle=0$ and $\langle \omega, [\delta,K_j]\rangle=0$(eq.(\ref{di1})). Hence $\langle L_{K_j} \omega,
\delta\rangle =0$ from the previous identity(\ref{liedr}). In other words $L_{K_j}\Omega_i$ is orthogonal to
$\Delta_{i+1}$. Since $\Delta_{i+1}$ is orthogonal to ${(\tilde{\Delta})_i}^*$ and $\Omega_i$, we have,
\begin{equation}
L_{K_j} \Omega_i \subseteq \Omega_i+{(\tilde{\Delta}_i)}^* \label{inf1}
\end{equation}
Now consider the same equation (\ref{liedr}), for all $\delta\in \tilde{\Delta}_i$ and all $\omega\in \Omega_i$ we
have $\langle \omega, \delta \rangle=0$. But since $[\delta,K_j]\notin \Delta_i$ we have $\langle \omega,
[\delta,K_j]\rangle\neq 0$ for some $\omega\in \Omega_i$, hence $\langle L_{K_j} \omega, \delta \rangle \neq 0$ as
well. Hence for any $\delta\in \tilde{\Delta}_i$ there exists an $\omega\in \Omega_i$ such that $\langle L_{K_j}
\omega, \delta \rangle \neq 0$. (i.e)
\begin{equation}
{(\tilde{\Delta}_i)}^* \subseteq L_{K_j}\Omega_i \label{inf2}
\end{equation}
Hence from eq.(\ref{inf1}) and (\ref{inf2}) we conclude that
\begin{equation}
{(\tilde{\Delta}_i)}^*+\Omega_i=\Omega_i+L_{K_j}\Omega_i, \forall 0\leq j\leq r.
\end{equation}
although it is possible to prove the stronger condition, $\Omega_i+\tilde{\Delta}^*_i=L_{K_j}\Omega_i$.
We now state the algorithm without proof:\\
\noindent
{\it Step 1:} Set $\Omega_0=\mbox{span}(dy(t,\xi))$.\\
{\it Step 2:} $\Omega_{i+1}=\Omega_i+L_{K_0}(\Omega_i)+\sum_{j=1}^r L_{K_i}(\Omega_i)$.\\
{\it Step 3:} The Algorithm converges to $\Omega^*=\Omega_i$ when $\Omega_{i+1}=\Omega_i,\forall i$.\\

Maximal invariant distribution $\Delta$ is such that $\Delta^*={\Omega^*}^\perp$. As seen in the proof each step
of Algorithm 1, removes a set from a vector which amounts to removing a finite dimension limited by dimension of
the tangent space. Hence the convergence of the algorithm is dependent on the finite dimensionality of the tangent
space at point $\xi(t)$ which can be guaranteed by the finiteness of control Lie Algebra, which will be studied in
the following sections.

\subsection{Observation space and Tangent Space}
In definition \ref{obs} the observation space spanned by $dy(t,\xi)$,$dL_{K_{i_0}}y(t,\xi)$,$\cdots$,
$dL_{K_{i_0}} \cdots L_{K_{i_n}} y(t,\xi)$, $\cdots \forall 0\leq i_0,\cdots, i_n \leq r \mbox{ and } n\geq 0 $
was defined and it can be easily seen that the necessary and sufficient condition for open loop decouplability
(\ref{invsp}) is equivalent to being orthogonal to the observation space according to def. \ref{obs}. The
orthogonality relation also follows from the simple Lie derivative identity,
\begin{equation}
L_{K_\tau}y(t,\xi)=\langle K_\tau, dy(t,\xi)\rangle
\end{equation}
From\cite{ganesan}, it can be seen that the one forms $dy(t,\xi)$,$dL_{K_{i_0}}y(t,\xi)$ etc can be expressed in
terms of the commutators of operators and hamiltonians, $C,H_0,H_i$. Infact the operations performed in the
observation space provide an alternative formulation to the theory developed in terms of the tangent space and
invariant distributions. As can be seen the structure of the output equation $y(t)=\langle
\xi(t)|C(t)|\xi(t)\rangle$ made possible the simplifications of Lie derivatives of scalar functions to commutators
of operators and enjoys ease of calculations when compared to computing Lie derivatives of vector and co-vector
fields, if one were to compute the invariant subspace. Hence it is to be noted that the necessary and sufficient
conditions for open loop decouplability can just be stated in terms of the observation space without ever having
to calculate the invariant distribution which is precisely what Theorem \ref{thm1} sets out to do. And it is also
to be noted that the Theorem is a consequence of the orthogonality relation in the observation space (Definition
\ref{obs}).

However when it comes to feedback decouplability the two different formalisms play equally important roles in
constructing a quantum system that might be decoupled using feedback. The observation space formalism provides
important necessary conditions ({\it in terms of the commutators of operators}) while designing a quantum control
system while the tangent space formalism is indispensable to calculating the feedback parameters
$\alpha(\xi(t)),\beta(\xi(t))$ once the system of interest is known to be decouplable using feedback.

\section{Synthesis of Feedback Parameters $\alpha(\xi), \beta(\xi)$}
In this section we study the explicit formulation of the feedback control that ensures complete decoupling of the
coherence functional from $H_{SB}$. It is to be seen that this formulation can be applied to outputs other than
the coherence functional we wish to monitor, like that of a non-demolition observable.
\begin{defn}
A distribution is said to controlled invariant on the analytic manifold $D_\omega$ if there exists a feedback pair
($\alpha, \beta$), $\alpha$, vector valued and $\beta$, matrix valued functions such that
\begin{align}
[\tilde{K}_0,\Delta ](\xi) \subset \Delta (\xi) \label{cid1}\\
[\tilde{K}_i,\Delta ](\xi) \subset \Delta (\xi) \label{cid2}
\end{align}
where,
\[
\tilde{K}_0 = K_0 + \sum_{j=1}^r \alpha_j K_j
\]
and
\[
\tilde{K}_i = \sum_{j=1}^r \beta_{ij}K_j
\]
\end{defn}
It is to be noted that $\tilde{K}_0$ and $\tilde{K}_i$ are the new drift and control vector fields of the control
system after application of feedback ($\alpha, \beta$). The problem of decoupling via feedback can now be cast in
the original framework of open loop decouplability by requiring that the feedback vector fields now satisfy the
open loop decouplability conditions viz.
\begin{align*}
[\tilde{K}_0,\Delta ](\xi) \subset \Delta (\xi)\\
[\tilde{K}_i,\Delta ](\xi) \subset \Delta (\xi)
\end{align*}
and that $\Delta$ be contained entirely within the null space of the output function (i.e),
\[
\Delta \subset \mbox{ker}(dy)
\]
With the above characterization of feedback decouplability the task now reduces to finding a distribution that
might satisfy the above invariance conditions with respect to the feedback vector fields, $(\tilde{K}_0,
\tilde{K}_1, \cdots , \tilde{K}_r)$, which in turn requires the knowledge of the feedback functions $\alpha$ and
$\beta$. What seems to be a deadlock situation can now be resolved by further simplifying the invariance condition
stated above.
\begin{lem} \label{feedbackthm}
An involutive distribution $\Delta$ defined on the analytic manifold $\mathcal{D}_\omega$ is invariant with
respect to the closed loop vector fields $(\tilde{K}_0, \tilde{K}_1, \cdots , \tilde{K}_r)$ for some suitable
feedback parameters $\alpha(\xi)$ and $\beta(\xi)$ if and only if,
\begin{align}
[K_0, \Delta] \subset \Delta + G \label{cid3}\\
[K_i, \Delta] \subset \Delta + G \label{cid4}
\end{align}
\end{lem}

Where $G$ is the distribution created by the control vector fields.
\begin{equation}
G=\mbox{ span } \{K_1, \cdots , K_r \}
\end{equation}
At this point it is possible to express the necessary and sufficient conditions for the feedback control system
$(\tilde{K}_0, \tilde{K}_1, \cdots , \tilde{K}_r)$ to be decoupled from the interaction vector field $K_I$ just as
we were able to provide conditions for open loop decouplability. Moreover the conditions can be expressed entirely
in terms of the open loop vector fields and the controlled invariant distribution without ever having to involve
the feedback parameters $\alpha(\xi)$ and $\beta(\xi)$. The following theorem provides the conditions,
\begin{thm}
The output $y(t,\xi)=\langle \xi | C(t) |\xi\rangle$ can be decoupled from interaction vector field $K_I$ via
suitable feedback $(\alpha,\beta)$ if and only if there exists an involutive distribution $\Delta$ such that,\\
\begin{align*}
[K_0, \Delta] \subset \Delta + G\\
[K_i, \Delta] \subset \Delta + G
\end{align*}
and $\Delta \subset \mbox{ker}(dy)$
\end{thm}
\begin{proof}
($\implies$) The following proof covers the lemma as well as the theorem above. Assuming that $\Delta$ is locally
controlled invariant or in other words invariant with respect to the closed loop vector fields $(\tilde{K}_0,
\tilde{K}_1, \cdots , \tilde{K}_r)$ for some feedback parameters $\alpha(\xi)$ and $\beta(\xi)$ within an open set
in $\mathcal{D}_\omega$. If $\tau\in \Delta$, then it can be seen that,
\begin{align*}
[\tilde{K}_i,\tau] &= [\beta_{ij}K_j,\tau] = \sum_{j=1}^r \beta_{ij}[K_j,\tau] - \sum_{j=1}^r (L_\tau
\beta_{ij})K_j
\end{align*}
as we know the left hand side is still contained within $\Delta$ and the last term on the right side is a linear
combination of vectors that generate $G$. Hence
\[
\sum_{j=1}^r \beta_{ij}[K_j,\tau] \in \Delta + G
\]
and since $\beta$ is assumed to be nonsingular it is possible to solve for individual $[K_j,\tau]$ by mere
inversion of the matrix $\beta_{ij}$ and can be found to be linear combination of vectors in $\Delta+G$ and hence,
\[
[K_i,\tau] \in \Delta + G
\]
Now consider,
\begin{align*}
[\tilde{K}_0,\tau]&=[K_0+\sum \alpha_j K_j, \tau]\\
&= [K_0,\tau] + \sum_{j=1}^r \alpha_j [K_j, \tau] - \sum_{j=1}^r (L_\tau \alpha_{j})K_j
\end{align*}
Since the left hand side belongs to $\Delta$ and since $[K_j,\tau]\in \Delta + G,1\leq j\leq r$ it can be
immediately seen that $[K_0,\tau] \in \Delta + G$ as well. \\

($\Longleftarrow$) For the proof of sufficiency the following geometric visualization is helpful. Let the
dimension of the distribution $\Delta$ be $d$. Since $\Delta$ is involutive there exist $d$ vectors fields,
locally {\it non-vanishing} in a neighborhood $U\subset\mathcal{S}_H\cap \mathcal{D}_\omega$ of $\xi$,
$\{|v_1\rangle,\cdots,|v_d\rangle\}\in T_\xi(\mathcal{M})$ that are linearly independent and,
\begin{equation}
\Delta = \mbox{span}\{|v_1\rangle,\cdots,|v_d\rangle\}
\end{equation}
s.t $[|v_i\rangle,|v_j\rangle]\in\Delta,\forall 1\leq i, j\leq d$. Now let the dimension of $\Delta+G$ at $\xi$ be
$d+q$. It is now possible to find another $q$ linearly independent vector fields labeled
$\{|v_{d+1}\rangle,\cdots,|v_{d+q}\rangle\}$, such that $[|v_i\rangle,|v_j\rangle]\in \Delta,\forall 1\leq i\leq
d, d+1\leq j \leq d+q$. As a special case one could think of a local co-ordinate basis that are mutually commuting
and linearly independent. Let the dimension of the tangent space at the point $\xi$ be $N$. Finally it is possible
to find $N-d-q$ additional linearly independent vectors that complete the vector space $T_\xi(\mathcal{M})$, by
Gram-Schmidt procedure or otherwise (i.e),
\begin{align}
T_\xi(\mathcal{M}) =
\mbox{span}\{&|v_1\rangle,\cdots,|v_d\rangle\,|v_{d+1}\rangle,\cdots,|v_{d+q}\rangle,\nonumber\\
&|v_{d+q+1}\rangle,\cdots,|v_N\rangle\}
\end{align}
It will be seen that the above requirement will be easily satisfied for the extension to control algebra to be
discussed following this proof. It is also to be noted that we haven't imposed any non-singularity restrictions on
the distributions above. Now the control vector fields $K_i\in G$ could be written as a linear combination of the
vector fields $\{|v_1\rangle,\cdots,|v_N\rangle\}$ at each point $\xi$.
\begin{align*}
&K_i=\sum_{j=1}^d c_{ij}|v_j\rangle+\sum_{j=d+1}^N c_{ij}|v_j\rangle, \forall 1\leq i \leq r\\
&K_i=K_i^d+K_i^o \{ \mbox{ where } K_i^d\in\Delta \mbox{ and } K_i^o\notin \Delta \}.
\end{align*}
The vector fields are devoid of components in $\Delta$. And since dimension of $\Delta+G$ is $d+q$ it can be seen
that the $r$ vectors $K_1^o,\cdots,K_r^o$ span a $q$ dimensional subspace. Hence it is always possible to generate
$q$ linearly independent vectors and $r-q$ zero vectors via suitable linear combinations of $K_1^o,\cdots,K_r^o$.
Let the linear combinations be such that,
\begin{align*}
&\sum_{j=1}^r\beta_{1j}K_j^o=|v_{d+1}\rangle + \sum_{j=d+q+1}^N \tilde{c}_{1j}|v_j\rangle\\
&\sum_{j=1}^r\beta_{2j}K_j^o=|v_{d+2}\rangle + \sum_{j=d+q+1}^N \tilde{c}_{2j}|v_j\rangle\\
&\vdots\\
&\sum_{j=1}^r\beta_{qj}K_j^o=|v_{d+q}\rangle + \sum_{j=d+q+1}^N \tilde{c}_{qj}|v_j\rangle\\
&\mbox{and}\\
&\sum_{j=1}^r\beta_{q+1,j}K_j^o=0\\
&\vdots\\
&\sum_{j=1}^r\beta_{r,j}K_j^o=0\\
\end{align*}
The $\beta_{ij}$ matrix so formed is precisely the feedback parameter that is used to generate the closed loop
vector fields $\tilde{K}_i,1\leq i \leq r$.
\[
\tilde{K}_i = \beta_{ij} K_j \mbox{ denoted by }\beta.K
\]
\begin{figure}
\begin{center}
\epsfysize=2.5in \epsfxsize=3.3in \epsffile{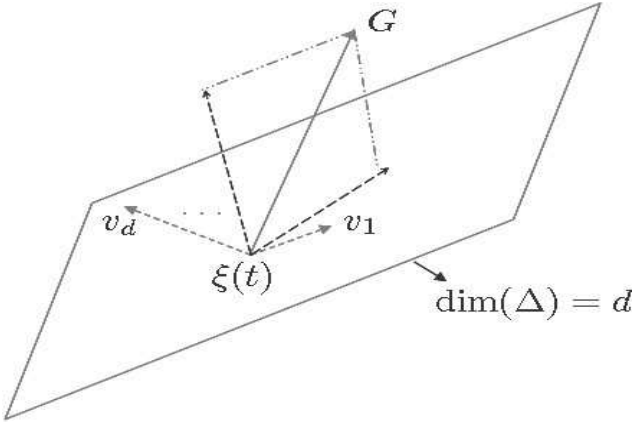}
\end{center}
\caption{The dimension of controlled invariant distribution is $d$ and the control distribution $G$ is partitioned
into $\{K_i^d\}$ and $\{K_i^o\}$. The basis vectors $|v_1\rangle,\cdots, |v_d\rangle$ span $\Delta$.}
\end{figure}
In order to prove this we note the action of the above linear combination on the open loop vector fields
$K_1,\cdots,K_r$,(i.e),
\begin{align}
&\sum_{j=1}^r\beta_{1j}K_j=\sum_{j=1}^d \tilde{c}_{1j}|v_j\rangle + |v_{d+1}\rangle + \sum_{j=d+q+1}^N \tilde{c}_{1j}|v_j\rangle \nonumber\\
&\sum_{j=1}^r\beta_{2j}K_j=\sum_{j=1}^d \tilde{c}_{1j}|v_j\rangle + |v_{d+2}\rangle + \sum_{j=d+q+1}^N \tilde{c}_{2j}|v_j\rangle \nonumber\\
&\vdots \nonumber\\
&\sum_{j=1}^r\beta_{qj}K_j=\sum_{j=1}^d \tilde{c}_{1j}|v_j\rangle + |v_{d+q}\rangle + \sum_{j=d+q+1}^N \tilde{c}_{qj}|v_j\rangle \label{eq1}\\
&\mbox{and}\nonumber\\
&\sum_{j=1}^r\beta_{q+1,j}K_i=\sum_{j=1}^d \tilde{c}_{1j}|v_j\rangle + 0\nonumber\\
&\vdots\nonumber\\
&\sum_{j=1}^r\beta_{r,j}K_j=\sum_{j=1}^d \tilde{c}_{1j}|v_j\rangle + 0 \nonumber
\end{align}
where the first terms on the right hand side of the above equations can be seen to be from $\beta K^d$ and the
later terms from $\beta K^o$. We will suppress the summation for ease of notation and all the following terms
below are assumed to be summations from $1,\cdots,r$ in the recurring index variable. Now from the necessary
conditions we have,
\begin{equation}
[\tau,K_j]\in \Delta+G, \forall \tau\in \Delta \mbox{ and } 1\leq j\leq r. \label{nesscond}
\end{equation}
and hence,
\begin{equation}
[\tau, \beta_{ij}K_j]= \beta_{ij}[\tau,K_j] + L_\tau(\beta_{ij})K_j \in \Delta + G. \label{tauktilde}
\end{equation}
and for $1\leq i \leq q$,
\begin{align*}
&[\tau, \beta_{ij}K_j]=[\tau,\sum_{j=1}^d \tilde{c}_{1j}|v_j\rangle + |v_{d+i}\rangle + \sum_{j=d+q+1}^N
\tilde{c}_{1j}|v_j\rangle]\\
&=[\tau,\sum_{j=1}^d \tilde{c}_{1j}|v_j\rangle]+[\tau,|v_{d+i}\rangle]+[\tau,\sum_{j=d+q+1}^N
\tilde{c}_{1j}|v_j\rangle]\\
\end{align*}
By noting that $\tau\in \Delta$, $\Delta$ is involutive and $|v_1\rangle,\cdots,|v_d\rangle$ commute with
$|v_{d+1}\rangle,\cdots,|v_N\rangle$, $[|v_i\rangle,|v_j\rangle]\in \Delta,\forall 1\leq i\leq d, d+1\leq j \leq
d+q$ the above equation can be seen to simplify to,
\[
[\tau, \sum_{j=1}^r \beta_{ij}K_j]\in \Delta+\sum_{j=d+q+1}^N L_\tau(\tilde{c}_{1j})|v_j\rangle
\]
but since we already have $[\tau,\beta_{ij}K_j]\in \Delta+G$ from (\ref{nesscond}) the above relation is possible
only if $L_\tau(\tilde{c}_{ij})=0$. Hence we have,
\[
[\tau, \sum_{j=1}^r \beta_{ij}K_j] = [\tau, \tilde{K}_i] \in \Delta.
\]
The argument is trivial for $q+1 \leq i \leq r$ and it can be easily seen that $[\tau,\sum_{1}^r \beta_{ij}K_j]\in
\Delta$ for all $1 \leq i \leq r$ and $\tau \in \Delta$. Now in order to construct the feedback parameter
$\alpha$, by an argument analogous to (\ref{tauktilde}), we can show that,
\[
[\tau,\tilde{K}_0]=[\tau,K_0+\sum_{i=1}^r \alpha_i K_i]\in \Delta+G
\]
because both $[\tau,K_0]$ and $[\tau,K_i]$ belong to $\Delta+G$. Let,
\begin{equation}
K_0 = \sum_{j=1}^d c_j |v_j\rangle + \sum_{j=d+1}^{d+q} c_j |v_j\rangle + \sum_{j=d+q+1}^{N} c_j |v_j\rangle
\label{k0}
\end{equation}
It is now possible to find a suitable linear combination of right hand side of equation set (\ref{eq1}) and the
above (\ref{k0}) in order to form $\tilde{K}_0$,
\[
\tilde{K}_0 = K_0+\tilde{\alpha}_i \tilde{K}_i = \sum_{j=1}^d \tilde{c}_j |v_j\rangle + \sum_{j=d+1}^{d+q} k_j
|v_j\rangle + \sum_{j=d+q+1}^N \tilde{c}_j |v_j\rangle
\]
where $k_j$'s are constants w.r.t $\xi$ and $t$ where as $\tilde{c}_j$'s are some functions of $\xi(t)$. In
particular by a suitable linear combination, $k_j$'s can all be made zero. It can again be seen that for all $\tau
\in \Delta$,
\[
[\tau,\tilde{K}_0]\in \Delta + \sum_{j=d+q+1}^N L_\tau (\tilde{c}_j) |v_j\rangle
\]
and hence $L_\tau (\tilde{c}_j)$ are equal to zero in order to satisfy the necessary conditions and hence,
\[
[\tau,\tilde{K}_0]\in \Delta
\]
The closed loop drift vector field was formed by setting $\tilde{K}_0$ equal to $K_0+\tilde{\alpha}.\beta.K$ for a
suitable row vector $\tilde{\alpha}$. Hence the feedback parameter $\alpha=\tilde{\alpha}.\beta$.
\end{proof}
In addition to proving the necessary and sufficient conditions we have also outlined a procedure to compute the
feedback parameters $\alpha(\xi)$ and $\beta(\xi)$ from the maximal controllability invariant distribution
$\Delta$, which elicits the application of Tangent space formalism in output decoupling. Hence it is imperative
that we compute the maximal invariant distribution $\Delta$ for the synthesis of feedback. From the necessary and
sufficient conditions we see that the distribution $\Delta$ has to satisfy conditions (\ref{cid1}-\ref{cid2}) or
equivalently (\ref{cid3}-\ref{cid4}) and that $\Delta\subset \mbox{ker}(dy)$ for complete decouplability.
Obviously (\ref{cid3}-\ref{cid4}) has the advantage that we do not need the knowledge of feedback parameters. Now,
similar to the open loop case we can formulate an algorithm in order to arrive at the much sought after invariant
distribution, the general idea being: Start out by assigning the whole of null space of $y(t)$ to $\Delta$ and
iteratively remove the part of the distribution that does not satisfy conditions (\ref{cid3}-\ref{cid4}).\\

\noindent
{\it Step 1:} Let $\Delta_0=\mbox{ker}(dy(t,\xi))$.\\
{\it Step 2:} $\Delta_{i+1}=\Delta_i-\{\delta\in\Delta_i:[\delta,K_j]\notin\Delta_i+G,\forall 0\leq j\leq r\}$\\
{\it Step 3:} Maximal invariant distribution is such that $\Delta^*=\Delta_{i+1}=\Delta_i$.\\

Employing the same logic as before in determining the open loop invariant distribution we can perform the
computation in the dual space $T^*_\xi(M)$ and arrive at the following algorithm which is easier to compute,\\

\noindent
{\it Step 1:} Let $\Omega_0=\mbox{span}(dy(t,\xi))$.\\
{\it Step 2:} $\Omega_{i+1}=\Omega_i+L_{K_0}(\Omega_i\cap G^\perp)+\sum_{j=1}^r L_{K_i}(\Omega_i \cap G^\perp)$.\\
{\it Step 3:} The Algorithm converges to $\Omega^*=\Omega_{i+1}=\Omega_i$.\\

\section{Extension to Control Algebra}
In the previous sections we provided a state feedback given by the vector $\alpha(\xi)$ and matrix $\beta(\xi)$
which were assumed to be analytical functions of the state $\xi$. In particular, the analyticity is required for
the proof of necessity as well as sufficient conditions. However, the class of analytic functions is too
restrictive in terms of feedback that can actually be implemented on the system. For example, by rapid pulses
which are arbitrarily strong and fast one can generate lie bracket of the vector control vector fields which can
act as a new control to the system available for feedback. In the light of non-analytic feedback it might be
necessary to modify the conditions that guarantee decouplability of the system. Another approach which is
sufficiently general would be to use the theory already developed for analytic feedback to systems whose control
vector fields belong to the control algebra of the original system,(i.e) we propose to use the system, where
$\hat{K}_i\in \{K_1,\cdots , K_r\}_{LA}=\mathcal{G}$. The theory of analytic feedback can now be extended to
controls from the control algebra instead of just the original set of controls. Hence we can restate the
conditions for decouplability in terms of the control algebra, which follows directly from the previous theorem
as,
\begin{lem}\label{controlalg}
The output $y(t)$ is decouplable via analytic feedback functions $\alpha(\xi)$ and $\beta(\xi)$ from the
interaction vector field $K_I$ if and only if there exists a controllability invariant distribution $\Delta$,
(i.e)
\begin{align}
[\Delta,\mathcal{G}]\subset \Delta \oplus \mathcal{G}\\
[\Delta,\mathcal{C}]\subset \Delta \oplus \mathcal{G}
\end{align}
where $\mathcal{C}=\{ad^j_{K_i}{K_0},i=1,\cdots,r;j=0,1\cdots\}$ and $\mathcal{G}=\{K_1,\cdots , K_r\}_{LA}$
\end{lem}
The above lemma just states a condition and does not provide an explicit formulation of the application of
feedback. In order to provide the analytic feedback we consider a modified system with additional control vector
fields generated from the original system. Consider the following modified system with finite dimensional control
algebra $\mathcal{G}$,
\begin{equation}
\frac{\partial \xi(t)}{\partial t} = K_0 |\xi(t)\rangle + \sum_{i=1}^{m} u_i \hat{K}_i |\xi(t)\rangle +
K_I|\xi(t)\rangle
\end{equation}
where the vector fields $\hat{K}_i \in \mathcal{G}$ which are generated by the vector fields of the original
system are such that $\mathcal{G} = \mbox{span}\{\hat{K}_1,\cdots,\hat{K}_m\}$, (i.e) the set of vector fields
$\hat{K}_i$, not necessary a linearly independent set form a vector space basis for $\mathcal{G}$. This is a
required condition as the analytic feedback functions which can only generate utmost linear combinations of the
existing control vector fields, (i.e) $\mbox{span}\{K_1,\cdots,K_r\}$ is inadequate to leverage the set of all
possible controls. Hence it is necessary to modify the original system in order to utilize the repertoire of all
possible controls for efficient feedback control. It is also to be noted that in so doing we do not alter the set
of reachable or controllable set of the original system, but altering the output decouplability instead which is
an observability property of the system.
\section{Examples}
As an example of the above formalism consider a single qubit and a two qubit system coupled to the environment,
\begin{align*}
\frac{\partial \xi(t)}{\partial t} &= \frac{\omega_0}{2}\sigma_z \xi(t) + \sum_k \omega_k b_k^\dagger b_k \xi(t) +
u_1 \sigma_x \xi(t) + u_2 \sigma_y \xi(t)\\
& + \sum_k \sigma_z(g_k b_k^\dagger + g_k^* b_k)\xi(t)
\end{align*}
with the output,
\[
y(t)=\langle \xi(t) |C| \xi(t)\rangle
\]

When we check against the necessary condition, $\sum_k \sigma_z(g_k b_k^\dagger + g_k^* b_k)\xi(t) \in
\mbox{ker}(dy(t))$ which we notice the single qubit system fails to satisfy, the conclusion that a single qubit
system is not decouplable coincides with results obtained earlier by operator algebra. Now, consider the following
two-qubit system eq.(\ref{2qubitcs})\,
\begin{align*}
&\frac{\partial |\xi(t)\rangle}{\partial t} &=& \left( \sum_{j=1}^2 \frac{\omega_0}{2}\sigma_z^{(j)} + \sum_k
\omega_k b_k^\dagger b_k\right)|\xi(t)\rangle \\
&&&+\sum_k \left(\sum_j \sigma_z^{(j)}\right)(g_k b_k^\dagger + g_k^* b_k) |\xi(t)\rangle \\
&+ (u_1(t) \sigma_x^{(1)} &+& u_2(t) \sigma_y^{(1)} + u_3(t) \sigma_x^{(2)} + u_4(t) \sigma_y^{(2)})|\xi(t)\rangle
\end{align*}
Which has a DFS of dimension 2, $\mbox{span}\{|01\rangle, |10\rangle\}$, the states within which remain coherent
in the absence of controls. The real problem arises in the presence of symmetry breaking perturbations or control
hamiltonians. Hence the problem at hand is to render the states coherent even in the presence of arbitrary
control. Consider the output of the form,
\[
y(t)=\langle \xi(t) |C| \xi(t)\rangle
\]

It can be clearly seen that the interaction vector field in deed belongs to $K_I = \sum_{j,k} \sigma_z^{(j)}(g_k
b_k^\dagger + g_k^* b_k) \xi(t) \in \mbox{ker}(dy(t)$, where $j=0,1$ and $k=0,1,\cdots $, but
\begin{align*}
[K_i,K_I]&=[\sigma_{x|y}^{(1)}\xi, \sum_j \sigma_z^{(j)}(g_k b_k^\dagger + g_k^* b_k) \xi]\\
&= c. \sum_k \sigma_{y|x}^{(1)}(g_k b_k^\dagger + g_k^* b_k)|\xi\rangle,\mbox{ eg. for }, i=1,2
\end{align*}
up to a constant $c$, neither belongs to the span of the control vector fields, control algebra generated by the
above vector fields or the controllability invariant distribution $\Delta$. The last condition can be seen by the
fact that $[K_i,K_I]$ does not belong to $\mbox{ker}(dy(t))$ and hence does not belong to $\Delta \subset
\mbox{ker}(dy(t))$ either. Now consider the two qubit system with bait, which was discussed in the earlier
section. The {\it control system} governing the mechanics following the Schr\"{o}dinger eq.(\ref{baitsysSE}) is
given by,
\begin{widetext}
\begin{align}
\frac{\partial |\xi(t)\rangle}{\partial t} =& \left( \sum_{j=1}^2 \frac{\omega_0}{2}\sigma_z^{(j)} + \sum_k
\omega_k b_k^\dagger b_k\right)\xi(t) + \sum_k \left(\sum_j \sigma_z^{(j)}\right)(g_k b_k^\dagger + g_k^*
b_k)\xi(t) + \left(u_1(t) \sigma_x^{(1)} + u_2(t) \sigma_y^{(1)} + u_3(t) \sigma_x^{(2)}\right. \nonumber \\
&\left.+ u_4(t) \sigma_y^{(2)} + \frac{\omega_0}{2}\sigma_z^{(b)} + u_5\sigma_x^{(b)}+u_6\sigma_y^{(b)}+u_7 J_1
\sigma_z^{(1)}\sigma_z^{(b)} + u_8 J_2 \sigma_z^{(2)}\sigma_z^{(b)}\right)\xi(t)+ u_9 \sum_k \sigma_z^{(b)}(w_k
b_k^\dagger + w_k^* b_k)\xi(t) \label{baitsys}
\end{align}
\end{widetext}
with $\sigma_{x|y|z}$now skew hermitian and the same output equation as before. It is seen that $K_I\in
\mbox{ker}(dy(t))$ and
\begin{align*}
[K_i,K_I]&=[\sigma_{x|y}^{(1)}\xi, \sum_j \sigma_z^{(j)}(g_k b_k^\dagger + g_k^* b_k) \xi] \\
&= c. \sum_k \sigma_{y|x}^{(1)}(g_k b_k^\dagger + g_k^* b_k) |\xi\rangle
\end{align*}
now belongs to the control algebra generated by the additional vector fields introduced by the {\it bait} system.
Hence the system which was designed in order to meet the necessary condition, $[\tilde{\mathcal{C}},H_{SB}]\subset
\tilde{\mathcal{C}}$, given by the observation space formalism is also seen to meet the conditions given by
tangent space or controllability invariant distribution formalism. A rather interesting scenario arises when the
drift vector field $K_0$ is a part of the ideal of $\mathcal{G}$ and the interaction vector field $K_I$ which is a
part of the invariant subspace $\Delta\subset \mbox{ker}(dy(t))$, is already contained within the control algebra,
(i.e) $K_I\in \mathcal{G}$. The necessary and sufficient conditions for decouplability using feedback are
trivially satisfied as $[K_I,\hat{K}_i]\in \mathcal{G} \forall \hat{K}_i\in \mathcal{G}$ and $[K_I,K_0]\in
\mathcal{G}$. Hence,
\begin{align}
[\Delta,\hat{K}_i]\subset \Delta \oplus \mathcal{G}\\
[\Delta,K_0]\subset \Delta \oplus \mathcal{G}
\end{align}
and the invariant subspace $\Delta$ can now be guaranteed to exist and at least one dimensional equal to
span$\{K_0\}$. Hence existence of feedback and decouplability is guaranteed for the above system.

\section{The Control System}
In the previous section we had only discussed a brief outline of the implementation of disturbance decoupling for
quantum systems. In this section we present the construction of actual control system and the control vector
fields. The bait qubit as discussed before was primarily used to get a {\it handle} on the environment so we may
generate vector fields that could help decouple the system from the vector field $K_I$. Let the following denote
the various hamiltonians acting on the system,
\begin{eqnarray}
&&H_0 =\sum_{j=1}^2 \frac{\omega_0}{2}\sigma_z^{(j)} + \sum_k \omega_k b_k^\dagger b_k,\nonumber\\
&&H_{SB} = \sum_k\left(\sum_j \sigma_z^{(j)}\right)(g_k b_k^\dagger + g_k^* b_k) \nonumber\\
&&H_1 = \sigma_x^{(1)}, H_2 = \sigma_y^{(1)}, H_3 = \sigma_x^{(2)}, H_4 = \sigma_y^{(2)} \nonumber \\
&&H_5 = \sigma_x^{(b)}, H_6 = \sigma_y^{(b)}, H_7 = J_1 \sigma_z^{(1)}\sigma_z^{(b)}, H_8 = J_2
\sigma_z^{(2)}\sigma_z^{(b)} \nonumber \\
&&H_9 = \sum_k \sigma_z^{(b)}(w_k b_k^\dagger + w_k^* b_k)
\end{eqnarray}
and let us denote by $K'_i$, the vector fields generated by the hamiltonian $H_i$, (i.e), $K_i=H_i|\psi\rangle$.
Now consider the particular back and forth maneuver via controls $u_6$ and $u_9$,
\begin{align*}
&u_6(\tau)=1, \mbox{ and } u_9(\tau)=0, \mbox{ for } \tau\in[0,t]\\
&u_6(\tau)=0, \mbox{ and } u_9(\tau)=1, \mbox{ for } \tau\in[t,2t]\\
&u_6(\tau)=-1, \mbox{ and } u_9(\tau)=0, \mbox{ for } \tau\in[2t,3t]\\
&u_6(\tau)=0, \mbox{ and } u_9(\tau)=-1, \mbox{ for } \tau\in[3t,4t]\\
\end{align*}
The corresponding unitary time evolution operator at the end of time instant $4t$ is given by,
\begin{align*}
U(4t) &= e^{(-iH_6 t)}e^{(-iH_9 t)}e^{(iH_6 t)}e^{(iH_9 t)}\\
&= \exp(-i[H_6,H_9]t^2+\mathcal{O}(t^3))
\end{align*}
the series expansion by Campbell-Baker-Hausdorff formula. In the limit that $t=dt\rightarrow0$. The effective
direction of evolution is given by the commutator of the corresponding hamiltonians, but to the second order in
time. Hence we could devise a control vector field in the direction given by the commutators of the corresponding
hamiltonians $H_6$ and $H_9$, where,
\[
[H_6,H_9] = c.\sigma_x^{(b)}\sum_k(w_k b_k^\dagger + w_k^* b_k)
\]
where $c$ is a {\it real} constant for a skew hermitian $H_6$ and $H_9$. In fact it is possible to generate any
direction of evolution with arbitrary strength corresponding to repeated commutators of the hamiltonians
$H_1\cdots H_9$ of the physical system (\ref{baitsys}). In order to compute commutators of tensor product
operators we use the following identitiy,
\begin{align*}
[A\otimes B, C\otimes D] = CA\otimes [B,D]+[A,C]\otimes BD
\end{align*}
With another control field $H_8$ entering the picture we could generate the following direction in conjunction
with the previous maneuver $[[H_8,H_5],[H_6,H_9]]$,
\begin{align}
&= c_1.[J_2\sigma_z^{(2)}\sigma_y^{(b)},\sigma_x^{(b)}\sum_k(w_k b_k^\dagger + w_k^* b_k)]\nonumber\\
&=c.\sigma_z^{(2)}\sigma_z^{(b)}\sum_k(w_k b_k^\dagger + w_k^* b_k) \label{comm1}
\end{align}
Consider the similar maneuver between controls $u_4,u_6 \mbox{ and }u_8$, which generates the direction of
evolution corresponding to the following repeated commutator,
\begin{align}
[H_4,H_8] &= [\sigma_y^{(2)},J_2\sigma_z^{(2)}\sigma_z^{(b)}] = c.\sigma_x^{(2)}\sigma_z^{(b)} \label{comm2}
\end{align}
where $c$ is a real constant for a skew hermitian $H_4, H_8$. Again, from operating on equations (\ref{comm1}) and
(\ref{comm2}) we get,
\begin{align}
&c_1[\sigma_x^{(2)}\sigma_z^{(b)}, \sigma_z^{(2)}\sigma_z^{(b)}\sum_k(w_k b_k^\dagger + w_k^*
b_k)]\nonumber\\
&=c_1.[\sigma_x^{(2)},\sigma_z^{(2)}].(\sigma_z^{(b)})^2.\sum_k(w_k b_k^\dagger + w_k^* b_k)\nonumber\\
&=c.\sigma_y^{(2)}.\mathbb{I}^{(b)}.\sum_k(w_k b_k^\dagger + w_k^* b_k)\label{comm3}
\end{align}
Hence we have generated an effective coupling between qubit 2 and the environment with the help of the bait qubit
and its interaction with the environment and qubit 2. It is important to note that the hamiltonian so obtained by
the above control maneuver now acts trivially on the hilbert space of the bait qubit, a property which will be
found to be extremely useful later. It is also possible to generate the $\sigma_x^{(2)}$ counterpart of the above
coupling by a similar maneuver, given by,
\begin{align}
c.\sigma_x^{(2)}.\mathbb{I}^{(b)}.\sum_k(w_k b_k^\dagger + w_k^* b_k)\label{comm4}
\end{align}
Again by a symmetric and totally similar argument we could generate a coupling between the environment and qubit
1, which would be given by,
\begin{align}
&c.\sigma_y^{(1)}.\mathbb{I}^{(b)}.\sum_k(w_k b_k^\dagger + w_k^* b_k) \mbox{ and } \label{comm5}\\
&c.\sigma_x^{(1)}.\mathbb{I}^{(b)}.\sum_k(w_k b_k^\dagger + w_k^* b_k)\label{comm6}
\end{align}
Now noting that the constants $c$ in the above equations could be controlled independently and arbitrarily, we can
write the preliminary form of the {\it actual} control system which achieves disturbance decoupling. Gathering
terms (\ref{comm3})-(\ref{comm6}), we construct the following control system for $\frac{\partial
|\xi(t)\rangle}{\partial t}$,
\begin{widetext}
\begin{align}
& = \left( \sum_{j=1}^2 \frac{\omega_0}{2}\sigma_z^{(j)} + \sum_k
\omega_k b_k^\dagger b_k\right)\xi(t) + \sum_k \left(\sum_j \sigma_z^{(j)}\right)(g_k b_k^\dagger + g_k^* b_k)
\xi(t) + (u_1\sigma_x^{(1)} + u_2\sigma_y^{(1)} + u_3\sigma_x^{(2)} + u_4\sigma_y^{(2)})\xi(t) \nonumber\\
&+\left(u_5\sigma_x^{(1)}\sum_k(w_k b_k^\dagger + w_k^* b_k)+ u_6\sigma_y^{(1)}\sum_k(w_k b_k^\dagger + w_k^* b_k)
+ u_7\sigma_x^{(2)}\sum_k(w_k b_k^\dagger + w_k^* b_k) + u_8\sigma_y^{(2)}\sum_k(w_k b_k^\dagger + w_k^*
b_k)\right)\xi(t)
\end{align}
\end{widetext}
By restructuring the control vector fields as above we are hoping to capture the entire control algebra by a
simple linear span of the control vector fields which is essential to analytical feedback theory. Let us again,
investigate the decouplability of the above control system from its necessary conditions, that $ (i) K_I\in \Delta
\subset \mbox{ker}(dy), (ii) [K_I, K_i]\in \Delta + G$, where $G=\mbox{span}(K_1\cdots,K_8)$, is the distribution
generated by the control vector fields above. By considering $[K_I,K_1]=\sigma_y^{(1)}\sum_k(w_k b_k^\dagger +
w_k^* b_k)|\xi(t)\rangle$, which is already contained within $G$. The conditions are also satisfied for the vector
fields $K_2,K_3$ and $K_4$. However with the vector field $K_5$, we note that $[K_I,K_5]$,
\begin{align*}
\tau = \left[\sum_k \left(\sum_j \sigma_z^{(j)}\right)(g_k b_k^\dagger + g_k^*
b_k)\xi(t),\right.\\
\left.\sigma_x^{(1)}\sum_k(w_k b_k^\dagger + w_k^* b_k)\xi(t)\right]\\
= c.\sigma_y^{(1)}\left(\sum_k(w_k b_k^\dagger + w_k^* b_k)\right)^2\xi(t)
\end{align*}
where w.l.o.g $w_k=c_1.g_k$ for an arbitrary constant $c_1\in\mathbb{C}$. For an infinite dimensional environment
the vector fields that contain higher powers of $\sum_k(w_k b_k^\dagger + w_k^* b_k)$, cannot be expressed as a
linear combination of its lower powers as can be seen from its action on a particular number state $|n\rangle$,
\begin{align*}
&(w_k b_k^\dagger + w_k^* b_k)|n\rangle &=& w_k \sqrt{n}|n-1\rangle + w_k^*\sqrt{n+1}|n+1\rangle\\
&(w_k b_k^\dagger + w_k^* b_k)^2|n\rangle &=& 2|w_k|^2 n|n\rangle + w_k^2 \sqrt{n(n-1)}|n-2\rangle\\
&&&+ {w_k^*}^2\sqrt{(n+1)(n+2)}|n+2\rangle
\end{align*}
for some $n$. In other words the above term is neither contained in $G$ nor in $\Delta \subset \mbox{ker}(dy)$,
because $L_\tau y(\xi,t)\neq 0$. The only way to correct the above situation is to include the vector $\tau$ as a
control vector field in the control system above. This can be achieved by similar maneuvers between the vector
fields above,(i.e), $\tau=c.[K_1,[K_1,H_2\xi]]$. Now again, since $\tau$ is a new control vector field, it must
satisfy condition $(ii)$ above. But $[K_I,\tau]= c.\sigma_x^{(1)}\left(\sum_k(w_k b_k^\dagger + w_k^*
b_k)\right)^3\xi(t)$, now generates the next higher power of the same environmental term, which necessitates us to
find a way to include that in our control vector fields as well. In fact, it is possible to generate any power of
the environmental term by repeated commutators, which is linearly independent of all the previous terms and hence
generates a new direction of flow within the analytic manifold. And it is impossible to include all the successive
powers in our control vector fields. Hence the best we could hope to achieve under the present circumstance is to
obtain an approximate solution to disturbance decoupling. It is to be noted that the above problem arises only in
an infinite dimensional environment and restricting the dimension of environment is a reasonably good
approximation. Hence we present a experimentally realizable scheme to demonstrate the theory of disturbance
decoupling to practical quantum systems. The following system captures the essence of the problem as well as the
solution itself. Before we present the example we summarize the results obtained thus far in a concise form. The
following table is helpful in noting the above decouplability results,

\begin{tabular}{|l|l|l|l|}
\hline &Open &Closed &Closed Loop\\
&Loop  &Loop &Restructured\\
\hline
Single Qubit &NO&NO&NO\\
\hline
Two Qubit &NO&NO&NO\\
\hline
Two Qubit or higher &NO&NO&$YES^*$\\
with bait qubit&&& \\
\hline
\end{tabular}

*-The system can be completely decoupled under the additional assumption of a finite dimensional
environment.

We note that the conditions for decouplability from Open loop to Closed loop to Closed Loop Restructured are
progressively relaxed. Hence a system that is not Closed Loop Restructured decouplable cannot be Closed Loop or
Open Loop decoupled.

{\bf Finite State Environment} Environment always appears to be in a stationary state(also called the Gibbs
State). An essential element of the stationary state which is most stable and extremely resilient is the coherent
state of an electromagnetic system. Coherent states is generated by the action of the displacement operator
$\hat{D}(\alpha)\equiv e^{(\alpha a^\dagger - \alpha^* a)}$ on the vacuum state $|0\rangle$. An electromagnetic
system when perturbed from one coherent state simply settles in another coherent state. It is labeled by a complex
number $\alpha$, that denotes the strength of the state. The state is given by,
\begin{equation}
|\alpha\rangle = e^{-1/2 |\alpha|^2}\sum_{n=0}^\infty \frac{\alpha^n}{\sqrt{n!}}|n\rangle
\end{equation}
where $|n\rangle$ is the number state. It can be the seen that the coefficients of higher $n$ decrease rapidly and
since squared sum of the coefficients is convergent, with major contribution from lower states it is a reasonable
approximation to neglect higher energy states of the electromagnetic system. In fact, this is the basis for the
experimental realization of ``dual rail optical photon quantum gates", where in only the $|0\rangle$ and
$|1\rangle$ photon states are used to represent the system under the premise that contributions from higher energy
photons are negligible. Hence we consider the following model for a finite state harmonic oscillator with $N$
energy states which will be later dubbed as the environment. The creation and annihilation operators act on the
system as follows,
\begin{align}
&a|n\rangle = \sqrt{n+1}|n+1\rangle \mbox{ for } n< N \nonumber\\
&a |n\rangle = 0 \mbox{ for } n\geq N \nonumber \\
&a^\dagger |n\rangle = \sqrt{n}|n-1\rangle \mbox{ for } n\leq N \mbox{ and } n> 0 \label{finiteenv}\\
&a^\dagger |n\rangle = 0 \mbox{ for } n> N \mbox{ and } n=0 \nonumber
\end{align}
It was recently shown by Fu et. al.\cite{schirmer3} in their model of {\it truncated} harmonic oscillator that
such a system up to energy state $N$ was feasible. Hence the schematic presented here can be readily implemented
if one were able to create and sustain a controllable interaction between the electromagnetic and spin system(the
bait). Now consider a single spin-1/2 system with hamiltonians $\sigma_z$, $\sigma_x$, and $\sigma_y$. The state
of the system is represented as $\psi = [c_0\mbox{ }c_1]'$ in the vector form where the coefficients correspond to
the two states and $c_0,c_1\in \mathbb{C} \mbox{ s.t } |c_0|^2+|c_1|^2=1$. The tangent vector to the system is
given by the action of the skew hermitian operators on the state $\psi$, (i.e), $\dot{\psi} = \sigma_{x|y|z}\psi =
[c_0'\mbox{ }c_1']'$ where $c_0',c_1' \in \mathbb{C}$.  In other words, in order to express any vector in the
tangent space as a real linear combination of other vectors we require at least 4 linearly independent vectors
given by $\sigma_z\psi,\sigma_x\psi,\sigma_y\psi,\mathbb{I}\psi$. For the case of a 2 spin-1/2 system the number
of linearly independent vectors required is 8, given by a subset of $\sigma_i\otimes \sigma_j \psi\mbox{ for
}\{i,j\}\in\{x,y,z,0\}$. For the case of 2 spin-1/2 system coupled to a 3 state environment, the tangent space is
$4\times 3$ dimensional and the number of linearly independent vectors required to span the entire tangent space
is $4\times 3\times 2 = 24$. In other words we require $24$ linearly independent control vector fields to make
absolutely sure that the conditions for decouplability are met. Let the environment be governed by a single 3
level harmonic oscillator. The different energy levels are given by $\{|0\rangle,|1\rangle,|2\rangle\}$, and a
general state in this basis is given by $|e\rangle = c_0|0\rangle + c_1|1\rangle + c_2|2\rangle$. We can now
examine the linearly independent vectors generated by the powers of the bath/environment operator for the three
level system by taking into account the defining relations (\ref{finiteenv}). As it can be seen that the following
6 vectors,
\begin{align*}
&\mathbb{I}\psi = c_0|0\rangle + c_1|1\rangle + c_2|2\rangle\\
&(w b^\dagger + w^* b)\psi =\\
& wc_1|0\rangle+w\sqrt{2}c_2|1\rangle+w^*c_0|1\rangle+w^*\sqrt{2}c_1|2\rangle \\
&(w b^\dagger + w^* b)^2\psi =\\
&[w^2(b^\dagger)^2+ww^* b^\dagger b + w^*w b b^\dagger + {w^*}^2 b^2]\psi\\
&(w b^\dagger + w^* b)^3\psi =[w^3(b^\dagger)^2 + w^2{w^*}(b^\dagger)^2 b + w^2{w^*}b^\dagger b b^\dagger\\
&+ w{w^*}^2 b^\dagger b^2 + w^2{w^*}b (b^\dagger)^2 + w{w^*}^2 b b^\dagger b + w{w^*}^2 b^2 b^\dagger]\psi\\
&(w b^\dagger + w^* b)^4\psi = [w^3w^*((b^\dagger)^2 b b^\dagger + b^\dagger b (b^\dagger)^2)\\
&+w^2{w^*}^2((b^\dagger)^2 b^2 + (b^\dagger b)^2 + b^\dagger b^2 b^\dagger+b(b^\dagger)^2 b + (b b^\dagger)^2\\
&+ b^2(b^\dagger)^2)+{w^*}^3w(b b^\dagger b^2+ b^2 b^\dagger
b)]\psi\\
&(w b^\dagger + w^* b)^5\psi =[w^3 {w^*}^2((b^\dagger)^2 b b^\dagger b + (b^\dagger)^2 b^2 b^\dagger + b^\dagger\\
& b(b^\dagger)^2 b + b^\dagger (b b^\dagger)^2 + b^\dagger b^2 (b^\dagger)^2 + b (b^\dagger)^2 b b^\dagger + b b^\dagger b (b^\dagger)^2)\\
&+ w^2{w^*}^3((b^\dagger b)^2 b + b^\dagger b^2 b^\dagger b + b (b^\dagger)^2 b^2 \\
&+ b(b^\dagger b)^2 + b b^\dagger b^2 b^\dagger + b^2(b^\dagger)^2 b+ b(b b^\dagger)^2)]\psi
\end{align*}
expressed in terms of the creation and annihilation operators of the bath, $b$ and $b^\dagger$ and do not contain
powers higher than 3 in their respective expansions and generate as many linearly independent vectors as possible
on $T_\xi(\mathcal{M})$, while operating on the state $\xi$. With the above linearly independent vectors we could
construct the new control system given by,
\begin{widetext}
\begin{align}
\frac{\partial |\xi(t)\rangle}{\partial t} =& \left( \sum_{j=1}^2 \frac{\omega_0}{2}\sigma_z^{(j)} + \sum_k
\omega_k b_k^\dagger b_k\right)|\xi(t)\rangle + \sum_k \sigma_z^{(j)}(g b^\dagger + g^* b) |\xi(t)\rangle +
\sum_{i=0}^5 u_{1i}\sigma_x^{(1)}(w b^\dagger + w^* b)^i|\xi(t)\rangle \nonumber\\
&+ \sum_{i=0}^5 u_{2i}\sigma_y^{(1)}(w b^\dagger + w^* b)^i|\xi(t)\rangle + \sum_{i=0}^5 u_{3i}\sigma_x^{(2)}(w
b^\dagger + w^* b)^i|\xi(t)\rangle + \sum_{i=0}^5 u_{4i}\sigma_y^{(2)}(w b^\dagger + w^* b)^i|\xi(t)\rangle
\end{align}
\end{widetext}
For the control system described above where the control vector fields $\{K_{ji}\}$, $0\leq i\leq 5$ and $1\leq j
\leq 4$, span the entire control algebra and hence,
\begin{equation}
[\Delta,K_{ji}]\subset \Delta + \mathcal{G}, 0\leq i \leq 5 \mbox{ and } 1\leq j \leq 4 \label{baitdelta}
\end{equation}

where $\mathcal{G} = \{K_1,\cdots, K_{24}\}_{LA} = \mbox{span}\{K_1,\cdots, K_{24}\}$. It now remains to know if
there exists a controlled invariant distribution $\Delta$, that satisfies the condition stated above. It can be
seen that since $\Delta$ is a subspace of the tangent space $T_\xi(\mathcal{M})$ at $\xi$, the equation above is
trivially satisfied. The only additional constraint that $\Delta$ needs to satisfy is that it be a part of the
$\mbox{ker}(dy)$, the nullspace of $y$ at the point $\xi$, which is a subspace of the tangent space
$T_\xi(\mathcal{M})$ itself. $\mbox{ker}(dy)$ is comprised of vectors of the form $H|\xi\rangle$ where $H$ is a
real linear combination(with coefficients possibly a function of the state $\xi$) of skew hermitian operators,
with the additional constraint that, $L_{(H\xi)}y=0$, which translates to the commutator, $[C,H]=0$. Since the
covector $dy$ is one dimensional for a scalar function $y$, the corresponding nullspace $\mbox{ker}(dy)$, would be
$n-1$ dimensional where $n$ is the dimension of the tangent space. Some of the vectors in $\mbox{ker}(dy)$ are,
\begin{align*}
&(\mathbb{I}^{(1)}\otimes\mathbb{I}^{(2)})(w b^\dagger + w^* b)^i|\xi(t)\rangle, 1\leq i \leq 5\\
&(\sigma_z^{(1)}+\sigma_z^{(2)})(w b^\dagger + w^* b)^i |\xi(t)\rangle, 1\leq i \leq 5\\
&i(\sigma_z^{(1)}+\sigma_z^{(2)})(w b^\dagger - w^* b)^i |\xi(t)\rangle, 1\leq i \leq 5  \mbox{ etc }
\end{align*}
where the operators $\sigma_z,\mathbb{I}$, above are to understood as skew hermitian operators as before. It is to
be noted that the algorithm presented in the previous section would terminate after the first iteration as the
condition is already satisfied and would yield $\mbox{ker}(dy)$ as $\Delta^*$, the maximal controlled invariant
distribution. The least value that $\Delta$ could take according to the necessary conditions of theorem
\ref{controlalg}, $K_I\in \Delta\subset \mbox{ker}(dy)$ is, the one dimensional vector space $\mbox{span}\{K_I\}$,
itself. The algorithm presented in the previous section is designed to yield the maximal invariant subspace, which
guarantees decouplability. But in order to compute the feedback we could work with any $\Delta$ that is a subspace
of maximal $\Delta^*$ and contains the minimal $\mbox{span}\{K_I\}$, as long as the condition (\ref{baitdelta}) is
satisfied.

{\bf Feedback Synthesis} In order to determine the feedback let us work with the minimal
$\Delta=\mbox{span}\{K_I\}$. It is possible to construct $n-1$ vectors where, $n=2\times
\mbox{dim}(T_\xi(\mathcal{M}))$ vectors $v_2,\cdots v_n\in T_\xi(\mathcal{M})$ that commutes with $v_1 = K_I$,
(i.e) $[v_1,v_j]=0$. Reindexing the control vector fields as $K_1 \cdots K_r$, where $r=n=24$ in this case, and
since $K_i$ span the tangent space we can write,
\begin{align}
v_j = \sum_{i=1}^r d_{ij} K_i
\end{align}
where $d$ is a non-singular real matrix. Hence we could rewrite,
\begin{align*}
\left(
\begin{array}{c}
K_1\\
K_2\\
\vdots\\
K_r
\end{array}
\right)=\left(
\begin{array}{ccc}
d_{11}&\cdots&d_{1r}\\
\vdots& &\vdots\\
d_{r1}&\cdots&d_{rr}
\end{array}
\right)^{-1} \left(
\begin{array}{c}
v_1\\
v_2\\
\vdots\\
v_r
\end{array}
\right)
\end{align*}
Following the proof of Theorem \ref{feedbackthm}, we can form the vectors, $K_i^o = S.v$, where, $S=d^{-1}$ but
with first column replaced by zeros,(i,e),
\begin{align*}
S=\left(
\begin{array}{cccc}
0&s_{12}&\cdots&s_{1r}\\
\vdots& &\cdots &\vdots\\
0&s_{r2}&\cdots&s_{rr}
\end{array}
\right)
\end{align*}
Now, the feedback parameter $\beta$ is such that
\begin{align*}
\beta \times \left(
\begin{array}{c}
K_1^o\\
K_2^o\\
\vdots\\
K_r^o
\end{array}
\right) = \left(
\begin{array}{c}
v_2\\
v_3\\
\vdots\\
v_r\\
0
\end{array}
\right)= \begin{array}{c} \underbrace{ \left(
\begin{array}{ccccc}
0&1&0&\cdots&0\\
0&0&1&0\cdots&0\\
\vdots& &\cdots& &\vdots\\
0&0&\cdots&0&1\\
0&0&\cdots&0&0
\end{array}
\right)}\\
J
\end{array}\left(
\begin{array}{c}
v_1\\
v_2\\
\vdots\\
v_r
\end{array}
\right)
\end{align*}
Since the above equations holds for all $v$ and $K$ vectors we could write $\beta(\xi)S=J$, but since the above
equation remains unaltered when $S$ is replaced with $d^{-1}$, we can calculate the feedback parameter as,
$\beta=J.d$. The closed loop vector fields are given by $\tilde{K}=\beta(\xi)K$. Similarly the parameter $\alpha$
can be calculated by incorporating $K_0$ in the equation. For any $K_0=\sum_{i=1}^r c_j v_j$, we can find an
$\tilde{\alpha}$ such that,
\begin{align}
K_0+\tilde{\alpha}_j \tilde{K}_j = c_1 v_1
\end{align}
for some $c_1$ as a function of the state $\xi$. The parameter $\alpha$ is given by $\alpha =
\tilde{\alpha}.\beta$ and the closed loop drift vector field is given by, $\tilde{K}_0=\sum_i \alpha_i.K_i$. It
can be seen that the above closed loop vector fields as in the proof does satisfy invariance w.r.t $\Delta$, (i.e)
$[\Delta,\tilde{K}_i]\subset \Delta, 0\leq i \leq r$. Hence the system is completely decoupled even in the
presence of symmetry breaking control hamiltonians via classical state feedback.

**we cannot find a suitable basis transformation using real matrices to a known set of commuting vectors such as $c_1|000\rangle, c_1|102\rangle,
\cdots\in T_\xi(\mathcal{M})$ etc where $c_1,c_2\in\mathbb{C}$, as performed in \cite{isidori}, where vectors were
transformed to coordinate basis in $\mathbb{R}^n$ in order to determine the feedback. Hence the task of finding
commuting vectors were simplified by such a transformation in the classical case. The difficulty is due to fact
that $(i)$ coefficients of the states complex and effectively carry twice the dimension, $(ii)$ tangent vectors at
point $\xi$ is different from that of another point $\xi_1$, hence a fixed coordinate transformation does not work
for every $\xi$. Whereas in the case of $\mathbb{R}^n$ tangent space at every point $x$ is the same.

It can also be noted the controllability properties of the system are unaltered in the presence of feedback. The
problem of disturbance decoupling is that of modifying the observability of the control system via feedback. It is
very well known from classical control theory that feedback can modify the observability properties of any system
but not the controllability properties. It is the observability of the decoherence that we intend to modify in the
above work by modeling it as a disturbance decoupling problem thus rendering the decoherence acting on the system
unobservable on the states of interest. However in order to accomplish the goals we had to introduce additional
couplings and a bait subsystem that were not a part of the system initially.
\section{Internal Model Principle}
In order to decouple the output from the environment one needs to determine the feedback coefficients
$\alpha(\xi)$ and $\beta(\xi)$ where both depend on the combined state of the system and environment. Hence one
needs to have a good estimate of the system as well as the environment itself for successful implementation of
feedback decoupling. In other words the state observer must include a model of the environment which would enable
us estimate its state. At this point, the important differences between classical and quantum decoupling problems
can be understood at the outset. The necessary condition in terms of the operator algebra
$[\tilde{\mathcal{C}},H_{SB}]\subset\tilde{\mathcal{C}}$ was instrumental in design of the bait subsystem. However
the structure of the system needed to be altered in order to,

$(i)$ Artificially induce coupling between qubits $1$, $2$ and the environment with the help of the bait.\\
$(ii)$ Generate vector fields in higher power of the environment operator to as to generate linearly independent
vectors.

Hence it was necessary to modify the core system in more ways than one in order to perform decoupling. Hence, even
though environment is an undesirable interaction the higher powers of the same helped us generate linearly
independent vectors in the tangent space, which was absolutely necessary for decoupling. Hence the environmental
coupling here befits the description of {\it necessary evil}. In classical dynamic feedback\cite{huangjie} the
design of controller depends on the exosystem. In contrast the state observer/estimator needs to know the model of
environment in order to estimate the combined state $\xi$ and calculate the feedback. Hence the model discussed
above could be thought of as the Internal Model Principle analog of quantum control systems. In addition classical
output regulation problem concerns with following a reference signal in the presence of environmental disturbace
that depends on a prescribed exosystem. On the other hand the disturbance decoupling problem focusses on
eliminating the effects of the environment.

%(iii) The model of the system as well as the environment along with a good estimate of the initial state is as
%good as having an observer to the system.

%(iv) Reinterpret the theorem from non-linear output reg. book. on IMP.

%(v) Classical internal model principle deals with perfect tracking and regulation in the presence of environmental
%disturbance, where as we are dealing with completely eliminating effects of the environment from our system.

\section{Bilinear input affine representation of Quantum Systems}
In this section we will attempt to highlight a few more important differences between the decoherence control in
quantum systems and disturbance decoupling of classical input affine systems in $\mathbbm{R}^n$.

(i) Classical noise is additive, $\dot{x} = f(x) + u_i g_i(x) + w p(x)$ and operate on the same vector space.
Whereas quantum noise is tensorial. The {\it noise} parameter $g_k$ and $g_k^*$ dictate the coupling between the
environment and the system , (i.e), $K_I = (\sigma_z^{(1)}+ \sigma_z^{(2)})\otimes(g_k^*
b_k+g_kb_k^\dagger)|\xi\rangle$ corresponds to the classical noise vector $p(x)$, and it can be easily seen that
there is no noise operating on the system in the classical sense. Hence decoherence is not classical noise.

(ii) Vector spaces in quantum control systems are over complex fields. This increases the dimensionality by 2 fold
in many instances where linearly combination has to be taken. Hence in order to generate every vector in a vector
space of $n$ independent states, we require $2n$ linearly independent vectors.

(iii) The necessary and sufficient conditions impose restrictions on the form of control hamiltonian that could
help decouple the system. From the conditions derived above, it is impossible to decouple one part of the system
from the other unless our control hamiltonians operate on the both the hilbert spaces non-trivially (i.e)
$H_i\in\mathcal{B}(\mathcal{H}_A\otimes\mathcal{H}_B)$, the set of linear operators in the joint hilbert space of
both the systems. It was in light of this condition that the bait system was originally introduced.

(iv) Distributions need not necessarily be singular. For instance the tangent space of an $\mathfrak{su}(2)$
system is spanned by $\sigma_z|\xi\rangle, \sigma_x|\xi\rangle, \sigma_y|\xi\rangle, \mathbb{I}|\xi\rangle$, where
$\xi=c_0|0\rangle + c_1|1\rangle$ and the operators are again assumed to be skew hermitian counterparts of
hermitian $\sigma_z,\sigma_x, \sigma_y$. Even though the four vectors are linearly independent for almost all
non-zero values of $c_0$ and $c_1$ the distribution is non-singular. Consider $|\xi\rangle = |0\rangle$ and the
corresponding tangent vectors are $-i|0\rangle, i|1\rangle, -|1\rangle, i|0\rangle$, whose real linear combination
is rank deficient. Hence it can be seen that the vector $|0\rangle$ {\it does not} belong to tangent space
$T_{|0\rangle}$ at the point $|\xi\rangle = |0\rangle$. In general the tangent vectors at point $\xi$ is different
from that of another point $\xi_1$. One of the most serious implications is that we cannot find a linear map that
transforms the distribution $\Delta$ to a constant $d$ dimensional distribution,
\begin{equation*}
T.\Delta = \left[ \begin{array}{c} I^{d\times d}\\0\end{array} \right]
\end{equation*}
at every point $\xi$, an approach that was used in Isidori\cite{isidori} to greatly simplify finding commuting
vectors $|v_1\rangle, \cdots |v_n\rangle$ in an $n$ dimensional tangent space. The commuting vectors were just
taken to be the co-ordinate basis at every point $x$.

\section{Conclusion}
In this work we provided the conditions and a step by step procedure to calculate a classical deterministic
feedback under which the 2-qubit system could be successfully decoupled from decoherence. As mentioned before the
analysis carried out in the bilinear form only helped us learn about the control hamiltonians helpful in
decoupling the system but also provided a solution under which the system would be completely decoupled as opposed
to partial or $n$th order decoupling discussed in various previous work. Such a control strategy would be
immensely helpful in performing decoherence free quantum computation thus enabling us to exploit the computational
speed up provided by quantum parallelism. However in order to determine the feedback one needs to have a good
estimate of the state of the system.


\begin{thebibliography}{999}
\baselineskip=15pt \footnotesize

\bibitem{opqusys}
H.-P. Breuer and F. Petruccione, The Theory of open quantum systems, {\it Oxford University Press}, 2002.

\bibitem{louisell}
William H. Louisell, Quantum Statistical Properties of Radiation, {\it John Wiley \& Sons, Inc}, 1973.

%\bibitem{sakurai}
%J. J. Sakurai, Modern Quantum Mechanics, {\it Addison-Wesley Publishing Company}, 1994.

\bibitem{qunet}
G. Mahler, V.A Weberru\ss, Quantum Networks, {\it Springer-Verlag}, 1998.

\bibitem{isidori}
Alberto Isidori, Nonlinear Control Systems, {\it Springer-Verlag}, 1995.

\bibitem{nijmeijer}
H. Nijmeijer, A. J. Van der Schaft, Nonlinear Dynamical Control Systems, {\it Springer-Verlag}, 1990.

\bibitem{mensky}
Michael B. Mensky, Quantum Measurements and Decoherence, Models and Phenomenology, {\it Kluwer Academic
Publishers}, 2000.

\bibitem{joos}
D. Giulini, E. Joos, C. Kiefer, J. Kupsch, I.-O. Stamatescu, H. D. Zeh, Decoherence and the Appearance of a
Classical World in Quantum Theory, {\it Springer}, 1996.

%\bibitem{chuang}
%Michael A. Nielsen and Isaac L. Chuang, Quantum Computation and Quantum Information, {\it Cambridge University
%Press}, 2000.

\bibitem{huangjie}
Jie Huang, Nonlinear Output Regulation, Theory and Application, {\it SIAM}, 2004.

\bibitem{itano}
W. M Itano, D. J. Heinzen, J. J. Bollinger and D. J. Wineland, Phys. Rev. A, {\bf 41}, pp 2295-2300, (1990).

\bibitem{zurek}
W. H. Zurek, Phys. Rev. D, {\bf 24}, 1516–-1525, 1981.

\bibitem{zurek1}
W. H. Zurek, ``Decoherence and the transition from quantum to classical - {\it Revisited}", {\it Los Alamos
Science}, No. 27, pp 2, 2002.

\bibitem{zurek2}
W. H. Zurek, Rev. of Modern Phyics, {\bf 75}(3), 715-775, 2003.

\bibitem{zurek3}
R. B-Kohout and W. H. Zurek, ``A Simple Example of ``Quantum Darwinism": Redundant information storage in
many-spin environments", arXiv:quant-ph/0408147.

\bibitem{ganesan}
N. Ganesan and T. J. Tarn, ``Control of decoherence in open quantum systems using feedback", {\it Proc. of the
$44^{th}$ IEEE CDC-ECC}, pp.427-433, Dec 2005.

\bibitem{ganesan1}  N. Ganesan, T. J. Tarn, ``Feedback Control of Decoherence by continuous measurements",
{\it Arxiv: quant-ph/0605044}, 2006.

\bibitem{buzek}
V. Bu\v{z}ek, G. Drobn\'{y}, R. Derka, G. Adam, and H. Widemann, ``Quantum State Reconstruction From Incomplete
Data", arXiv:quant-ph/9805020.

\bibitem{htc}
G. M. Huang, T. J. Tarn, J. W. Clark, ``On the controllbility of quantum mechanical systems", {\it J. Math.
Phys},24(11), pp 2608, Nov 1983.

%\bibitem{kunita1}
%H. Kunita, {\it Proc. Int. Sym. on SDE}, p. 163, 1976.
%
%\bibitem{kunita2}
%H. Kunita, ``On the controllability of nonlinear systems with applications to polynomial systems", {\it Appl.
%Math. Optm.}, 5, 89, 1979.
%
%\bibitem{nelson}
%E. Nelson, ``Analytic Vectors", {\it Ann. Math}, 70, 572, 1959.
%
%\bibitem{sussman}
%H. J. Sussman and V. Jurdjevic, ``Controllability of nonlinear systems", {\it J. Diff Eqns}, 12, 95, 1972.
%
%\bibitem{chow}
%W. L. Chow, ``Uber Systeme von linearen partiellen Differentialgleichungen erster Ordnung", {\it Math. Ann.}, 117,
%98, 1940.
%
%\bibitem{dong1}
%S-H Dong, Y Tang, G-H Sun, F Lara-Rosano, M Lozada-Cassou, ``Controllability of pure states for the
%P\"{o}schl-Teller potential with a dynamical group SU(2)", {\it Annals of Phys.}, {\bf 315}, 566, 2005.

%\bibitem{schirmer1}
%S G Schirmer, A I Solomon and J V Leahy, ``Degrees of controllability for quantum systems and application to
%atomic systems", {\it J. Phys. A: Math. Gen.}, {\bf 35}, 4125, 2002.
%
%\bibitem{schirmer2}
%S G Schirmer, A I Solomon and J V Leahy, ``Criteria for reachability of quantum states", {\it J. Phys. A: Math.
%Gen.}, {\bf 35}, 8551, 2002.

\bibitem{schirmer3}
H. Fu, S.G. Schirmer and A.I. Solomon, ``Complete controllabi1ity of finite-level quantum systems", {\it J. Phy.
A:Math. and Gen.} , {\bf 34}(8), pp. 1679-1690, 2001.

\bibitem{brockett}
R. W. Brockett, C. Rangan, A. M. Bloch, ``The controllability of infinite quantum systems", {\it Proc. of the
$42^{nd}$ IEEE CDC}, {\bf 1}, pp. 428-433, 2003.

%\bibitem{dong2}
%S-H Dong, Y Tang, G-H Sun, ``On the controllability of a quantum system for the Morse potential with compact group
%SU(2)", {\it Phys. Lett. A}, {\bf 320}, 145, 2003.

\bibitem{brune1}
M. Brune, E. Hagley, J. Dreyer, X. Ma\^{i}tre, A. Maali, C. Wunderlich, J. M. Raimond, and S. Haroche, Phys. Rev.
Lett., {\bf 77}(24), 4887, 1996.

\bibitem{brune2}
S. Haroche, M. Brune and J. M. Raimond, Phil. Trans. R. Soc. Lond. A, {\bf 355}, 2367-2380, 1997.

\bibitem{brune3}
J. M. Raimond, M. Brune and S. Haroche, ``Reversible Decoherence of Mesoscopic Superposition of Field States",
Phys. Rev. Lett., {\bf 79}(11), pp 1964-1967, 1997.

\bibitem{geremia}
J. M. Geremia, J. Stockton, H. Mabuchi, arXiv:quant-ph/0401107{\bf v4}

\bibitem{wiseman}
H. M. Wiseman, Phys. Rev. A,{\bf 49}(3), pp 2133-2150, 1994.

%\bibitem{uchiyama}
%C. Uchiyama, M. Aihara, ``Multipulse control of decoherence", {\it Phys. Rev. A}, 66, 032313, 2002.
%
%\bibitem{horoshko}
%D. B. Horoshko and S. Y. Kilin, ``Decoherence slowing via feedback", {\it Journal of Modern Optics}, 44(11/12), p
%2043, 1997.

\bibitem{lidar}
D. A. Lidar, I. L. Chuang and K. B. Whaley, ``Decoherence-free subspaces for quantum computation", {\it Phys. Rev.
Letters}, 81(12), p 2594, 1998.

\bibitem{lidar2}
D. Lidar, L.-A. Wu, ``Reducing Constraints on Quantum Computer Design by Encoded Selective Recoupling", {\it Phys.
Rev. Lett.}, 88, 017905, 2002.

\bibitem{lidar3}
D. Lidar, L.-A. Wu, ``Encoded recoupling and decoupling: An alternative to quantum error-correcting codes applied
to trapped-ion quantum computation", {\it Phys. Rev. A}, 67, 032313 (2003).

%\bibitem{viola1}
%L. Viola, E. Knill and S. Lloyd, ``Dynamical decoupling of open quantum systems", {\it Phys. Rev. Letters},
%82(12), p 2417, 1999.
%
%\bibitem{viola2}
%L. Viola, S. Lloyd and E. Knill, Phys. Rev. Lett., {\bf 83}(23), 4888, 1999.
%
%\bibitem{viola3}
%L. Viola, ``Quantum control via encoded dynamical decoupling", {\it Phys. Rev. A}, 66, 012307, 2002.

\bibitem{doherty}
A. C. Doherty, K. Jacobs and G. Jungman, ``Information, disturbance and Hamiltonian feedback control", {\it Phys.
Rev. A}, 63, 062306, 2001.

\bibitem{fortunato}
E. M. Fortunato, L. Viola, J. Hodges, G. Teklamariam and D. G. Cory, ``Implementation of Universal Control on a
Decoherence-Free Qubit", {\it New Journal of Phys.}, {\bf 4}, 5.1-5.20, 2002.

\bibitem{mohseni}
M. Mohseni, J. S. Lundeen, K. J. Resch and A. M. Steinberg, ``Experimental application of Decoherence-Free
Subspaces in an Optical Quantum Computing Algorithm", {\it Phys. Rev. Lett.}, {\bf 91}(18), 187903, 2003.

\bibitem{ollerenshaw}
J. E. Ollerenshaw, D. A. Lidar and L. E. Kay, ``Magnetic Resonance Realization of Decoherence-Free Computation",
{\bf 91}(21), 217904, 2003.

\bibitem{kielpinski}
D. Kielpinski, C. Monroe and D. J. Wineland, ``Architecture for a large-scale ion-trap quantum computer", {\it
Nature}, 417, pp 709-711 , 2002.

\bibitem{brown}
K. Brown, J. Vala, and K. B. Whaley, ``Scalable ion trap quantum computation in decoherence-free subspaces with
pairwise interactions only", {\it Phys. Rev. A} {\bf 67}, 012309, 2003.

\bibitem{xue}
F. Xue et. al, Phys. Rev. A,{\bf 73}, 013403, 2006.

\bibitem{roa}
L. Roa and A. Delgado et. al, Phys. Rev. A,{\bf 73}, 012322, 2006.

\bibitem{wallentowitz}
S. Wallentowitz, ``Quantum theory of feedback of bosonic gases", {\it Phys. Rev. A}, 66, 032114, 2002.

\bibitem{buscemi}
F. Buscemi, G. Chiribella, and G. M. D'Ariano, Phys. Rev. Lett., {\bf 95}, 090501, 2005.

\bibitem{altifini}
C. Altifini, J. Math. Phys, {\bf 44}(6), pp 2357-2372, 2003.

\bibitem{jacobs}
K. Jacobs, ``How to project qubits faster using quantum feedback", {\it Phys. Rev. A}, 67, 030301(R), 2003.

\bibitem{omnes}
R. Omn\'{e}s, ``General theory of the decoherence effect in quantum mechanics", {\it Phys. Rev. A}, 56(5), pp
3383, 1997.

\bibitem{proto}
V. Protopopescu, R. Perez, C. D'Helon and J. Schmulen, ``Robust control of decoherence in realistic one-qubit
quantum gates", {\it J. Phys A:Math. Gen.,} 36, pp 2175, 2003.

%\bibitem{barginsky}
%V. B. Barginsky, Y. I. Vorontsov, K. S. Thorne, ``Quantum nondemolition measurements", {\it Science}, Vol 209, No.
%4456, pp 547, 1980.
%
%\bibitem{caves}
%C. M. Caves, K. S. Thorne, R. W. P. Drewer, V. D. Sandberg and M. Zimmerman, ``On the measurement of a weak
%classical force coupled to a quantum-mechanical oscillator", {\it Rev. of Mod. Phys.}, 52(2), Part I, pp 341,
%1980.

\bibitem{shor}
P. Shor, ``Scheme for reducing decoherence in quantum computer memory", {\it Phys. Rev. A}, 52, 2493, 1995

\bibitem{calderbank}
A. R. Calderbank and P. W. Shor, ``Good quantum error-correcting codes exist", {\it Phys. Rev. A}, 54, 1098, 1996.

\bibitem{wallentowitz}
S. Wallentowitz, ``Quantum theory of feedback of bosonic gases", {\it Phys. Rev. A}, 66, 032114, 2002.

%\bibitem{ong1}
%J. W. Clark, C. K. Ong, T. J. Tarn and G. M. Huang, ``Invertibility of quantum-mechanical control systems", {\it
%Math. Systems Theory}, 17, pp 335, 1984.
%
%\bibitem{ong2}
%J. W. Clark, C. K. Ong, T. J. Tarn and G. M. Huang, ``Quantum nondemolition filters", {\it Math. Systems Theory},
%18, pp 33, 1985.

\bibitem{lan}
C. Lan, T. J. Tarn, Q. S. Chi and J. W. Clark, ``Analytic controllability of time-dependent quantum control
systems", {\it J. Math. Phys}, April 2005.

\bibitem{dayawansa}
W. Dayawansa, D. Cheng, W. M. Boothby and T. J. Tarn, ``Global $(f,g)$-invariance of nonlinear systems", {\it SIAM
J. Control and Optimization}, Vol 26, No. 5, pp 1119, 1988.

\end{thebibliography}
\end{document}